\documentclass[preprint, 12pt]{elsarticle}
\usepackage[a4paper,
bindingoffset=0.2in,
left=1in,
right=1in,
top=0.75in,
bottom=1in,
footskip=.25in]{geometry}
\usepackage{amssymb}
\usepackage{amsmath}
\usepackage{url}
\usepackage{graphicx}
\usepackage{fonttable}
\usepackage{rotating}
\usepackage{ulem}
\usepackage{etoolbox}{}
\usepackage{braket}
\usepackage{epstopdf}
\usepackage{caption}
\usepackage{multirow}
\usepackage{color}

\def\be{\begin{equation}}
\def\ee{\end{equation}}
\def\bea{\begin{eqnarray}}
\def\eea{\end{eqnarray}}
\def\bdm{\begin{displaymath}}
\def\edm{\end{displaymath}}

\journal{EPJA}

\begin{document}
	
	\begin{frontmatter}
		
		\title{Allowed and forbidden $\beta$-decay log $ft$ values of neutron-rich Pb and Bi isotopes}

		\author {Necla \c{C}akmak$^{1}$}
		\author {Jameel-Un Nabi$^{2}$}

		\author {Arslan Mehmood$^{2}$}
		\author{Asim Ullah$^{3}$}
		\author {Rubba Tahir$^{2}$}

		\address{$^1$University of Karab\"uk, Department of Physics, 78050 Karab\"uk, T\"urkiye}
		\address{$^2$University of Wah, Quaid Avenue, Wah Cantt 47040, Punjab, Pakistan}
		\address{$^3$ University of Swabi, Khyber Pakhtunkhwa, Pakistan}
		
\begin{abstract}
			
The $\beta$-decay log $ft$ values for $^{210-215}$Pb \textrightarrow $^{210-215}$Bi and $^{210-215}$Bi \textrightarrow $^{210-215}$Po transitions in the north-east region of $^{208}$Pb nuclei are estimated  using the proton-neutron quasiparticle random phase approximation model. The pn-QRPA equations were solved using the schematic model approach. The Woods-Saxon (WS)  potential was inserted as a mean-field basis and nuclei were treated as spherical. Allowed Gamow-Teller (GT) and first-forbidden (FF) transitions were investigated in the particle-hole (\textit{ph}) channel. The calculated log $ft$ values of the allowed GT and FF transitions using the pn-QRPA(WS) were found  closer to the experimental values.  Later we performed calculation of $\beta$-decay rates in stellar environment. Here we solved the RPA equations in deformed Nilsson basis, both in the  particle-particle (\textit{pp}) and particle-hole (\textit{ph}) channels. Allowed $\beta$-decay  and unique first-forbidden (U1F) rates were calculated in stellar matter.  For certain cases, the calculated U1F contribution was much more than the allowed $\beta$-decay rates under prevailing stellar conditions, in line with previous findings. Increasing  temperature of the stellar core affected the allowed GT rates more than the U1F rates.
\end{abstract}
		
\end{frontmatter}
	
\section{Introduction}
\label{intro} The supernova explosion provides useful information  for investigating the evolution and death of massive  stars.  The lepton-to-baryon fraction of the stellar matter is controlled by the weak interaction rates. These include $\beta^{\pm}$ decay and $\beta^{\pm}$ capture rates~\cite{Ful80}.   The weak interaction reactions contribute to the collapse of massive stars. Charge-changing transitions provide reliable estimate of the stellar core and  useful information about the strength of shock wave produced by supernova outburst \cite{Bor06}. The charge-changing transitions are governed by  Gamow-Teller (GT) and Fermi operators. As one moves away from the stability line, first-forbidden (FF) transitions contribute significantly. The FF $\beta$ decay rates are of key importance due to their enlarged phase space in neutron-rich nuclei \cite{War88}. 

The core-collapse supernova event is preferable cite for the formation of heavier elements in the universe ($r$-process)~\cite{Bur57}. The $r$-process is useful in examining the  $\beta$ decay parameters near the neutron drip line \cite{Bur57}. The $\beta$ decay properties of neutron rich nuclei are impotent to understand the $r$-process. The actual site of the astrophysical $r$-process is still uncertain because it occurs mostly in an explosive environment with relatively high temperature $(T \sim 10^9 K)$ and high neutron density $(> 10^{20} g/cm^{-3})$ \cite{Kra93,Cow91,Woo94,Wan06}. The $r$-process contributes significantly to heavy element  nucleosynthesis. Probable cites for $r$-process include core-collapse supernovae \cite{Thi17} and neutron star mergers \cite{Suz12}. Authors in ref.~\cite{Bur57} hinted that heavy element synthesis for neutron-deficient isotopes cannot be built by either the $s$- or the $r$-process. They argued that the $p$-process was responsible  for the formation of proton-rich isotopes of a large number of heavy elements. The $p$-process includes ($p, \gamma$) and possibly ($\gamma , n$) reactions on material which has already been synthesized by the $s$- and $r$-processes. The pre-requisites for the $p$-process, as suggested by ref.~\cite{Bur57} are core density around $[> 10^{2} g/cm^{-3}]$, a hydrogen rich environment and a high temperature [$(2-3) \times 10^9 K]$. These conditions are roughly replicated in the envelope of a supernova of Type II. Thus $\beta$-decay observables of neutron-deficient isotopes are also important for a better understanding of the heavy element nucleosynthesis (see also ref.~\cite{Arn76}).

Despite the challenges posed in half-life calculations of nuclei near N = 126,  researchers identified a new $\beta$-decaying isomeric state  $(8^-)$ in $^{214}$Bi by using the $\gamma$ ray spectroscopy at the ISOLDE Decay Station at the CERN-ISOLDE facility~\cite{And21}.   Different parameters such as log $ft$, $\beta$-decay feeding fraction were reported. The authors performed shell model calculations to validate the new $\beta$-decaying state as an isomer and assigned shell-model levels to several high-spin states in $^{214}$Po.  Later, a shell-model study of the log $ft$ values for the forbidden $\beta$-decay transitions was performed around the lead region~\cite{Sha22}. The calculation employed the KHPE interaction in valence shell for Z = (82 - 126) and N = (126 - 184). The half-lives and $Q$-value calculations were performed without any truncation. 

%There are many considerable theoretical approaches that can be used to calculate the weak decay properties. First approach is the macroscopic model; i.e the Gross theory of beta decay \cite{Pfe02,Mol03}, and the second approach is the microscopic shell model; which differs in the way it explains the initial ground to final excited states. The third approach is the semi-microscopic approaches such as the Kratz Herrmann \cite{Bor06}; the quasi-particle random-phase approximation (QRPA); the finite range droplet method (FDRM) \cite{Bur57,Eng99}, the density function (DFT) theory with QRPA method \cite{Hal67}, and the Hartree-Fock Bogoliubov theory \cite{Ken05}. 
%$\beta$-decay properties  were studied using the pn-QRPA model with a separable multi-shell interaction on top of axially symmetric-deformed mean-field calculations with an axially deformed Nilsson potential~\cite{Sta90,Hir93,Hom96}. The model possessed excellent predictive power (see Table A of refs.~\cite{Sta90,Hir93} and Tables (III - VIII) of ref.~\cite{Hom96}). The same pn-QRPA model was used to calculate the stellar rates, including highly unstable nuclei in ref.~\cite{Nab99,Nab99a,Nab04}. Later, the model was used to examine (U1F) transitions ($\Delta J = 2$) under stellar conditions for neutron-rich nickel~\cite{Nab14,Nab16}, \textbf{germanium, zinc}~\cite{Nab15} and copper isotopes~\cite{Maj17}. The calculations were also performed for heavy odd-odd and even-even nuclei in the mass range 70$\le$ A $\le$ 214 including U1F rates \cite{Nab17}. 
Motivated by the recent measurement~\cite{And21} and shell-model calculations~\cite{And21,Sha22}, we report our investigation of $\beta$-decay observable of neutron-rich Pb and Bi isotopes, using the (pn-QRPA) model. Halbleib and Sorensen \cite{Hal67} developed the pn-QRPA model by simplifying the random phase approximation (RPA). In the present work, we used two distinct versions of  the pn-QRPA model to calculate $\beta$-decay properties. These include log $ft$ values, partial \& total half-lives and stellar rates of $^{210-215}$Pb and $^{210-215}$Bi. Our calculations include both allowed GT and FF transitions. The terrestrial calculations were performed assuming the nuclei to be spherical. Single-particle energies were calculated using the Woods-Saxon potential and calculations were performed in the particle-hole channel. This model is referred to as the pn-QRPA(WS). A theoretical description of the charge-exchange spin-dipole transitions using the Pyatov’s restoration method was described earlier for the chosen model~\cite{Unlu23}. Employing the restoration procedures, rendered the mathematical formalism free of effective interaction strength parameters. Later, we introduced the nuclear deformation in our investigation and computed stellar rates for $^{210-215}$Pb and $^{210-215}$Bi. This pn-QRPA model computed single-particle energies in a deformed Nilsson basis and is referred to as pn-QRPA(N). The pn-QRPA(N) calculations were performed in both particle-particle and particle-hole channels. 
Essentials of theoretical formalism of both models is explained in section 2. Results and discussion are presented in section 3. Finally, we conclude our findings in section 4.    
   
	\section{Formalism}
	\label{sec:2}
	
	Two pn-QRPA models using different single-particle energies and solutions were considered for the current investigation. For terrestrial conditions, we used the pn-QRPA(WS) model for calculation of allowed and FF transitions. For stellar rates, we used the pn-QRPA(N) model and  calculated allowed and unique first forbidden transitions in stellar matter. The formalism for the two pn-QRPA models is briefly discussed in the following subsections.
	\subsection{The pn-QRPA(WS) Model}
The log$ft$ and half-lives values for GT and FF transitions were calculated employing a spherical schematic model (SSM). The Woods-Saxon (WS) potential was used to calculate the single-particle energies and wave functions. The calculation was performed in the particle-hole (\textit{ph}) channel to calculate the eigenvalues and eigenfunctions of the Hamiltonian. The Gamow-Teller ($J^{\pi}= 1^{+}$) and first forbidden ($J^{\pi}= 0^{-},1^{-},2^{-}$) excitations in odd-odd nuclei were generated from the correlated ground state of the parent nuclei through the charge-exchange spin-spin and spin-dipole interactions. The eigenstates of the single quasiparticle Hamiltonian $H_{sqp}$ was later used as a basis. The Hamiltonian of the allowed and first forbidden transitions was chosen as

\begin{eqnarray}
	H_{SSM}=H_{sqp}+h_{ph},
\end{eqnarray}
where $H_{sqp}$ is the single quasiparticle Hamiltonian. The effective interaction was considered in the \textit{ph} channel, for both GT and FF transitions, and denoted by $h_{ph}$. The single quasi-particle Hamiltonian of the system was given by
\begin{eqnarray}
	{H}_{sqp}=\sum_{j_\tau m_\tau}\varepsilon_{j_\tau m_\tau}{\hat{\alpha}}_{j_pm_p}^\dag {\hat{\alpha}}_{j_nm_n},~~~~(\tau=n,p)
\end{eqnarray}
where $\varepsilon_{j_\tau m_\tau}$ is the single quasi-particle energy of the nucleons with angular momentum $j_\tau m_\tau$. The ${\hat{\alpha}}_{j_pm_p}^\dag$ and ${\hat{\alpha}}_{j_nm_n}$ are one quasi-particle creation and annihilation operators, respectively.

The charge-exchange spin-spin and spin-dipole effective residual interactions were determined using 
\begin{eqnarray}
	h_{ph}=\frac{2\chi_{ph}}{g_{A}^2}\sum_{j_{\tau}j_{\tau'}\mu\mu'}[b_{j_{p}j_{n}}(\lambda)A^{\dag}_{j_{p}j_{n}}(\lambda\mu)+(-1)^{(\lambda+\mu)}\bar{b}_{j_{p}j_{n}}(\lambda)A_{j_{p}j_{n}}(\lambda-\mu)] \nonumber
\end{eqnarray}
\begin{eqnarray}
	\times[b_{j_{p'}j_{n'}}(\lambda)A_{j_{p'}j_{n'}}(\lambda\mu')+\bar{b}_{j_{p'}j_{n'}}(\lambda)A^{\dag}_{j_{p'}j_{n'}}(\lambda-\mu'],
\end{eqnarray}
where ${\chi}_{ph}$ represents the particle-hole (\textit{ph}) effective interaction constant chosen as
$$
{\chi}_{ph}=\begin{cases} 
	5.2A^{0.7}\mathrm MeV         		  &  \text{GT}~~1^{+} \\ 
	30A^{-5/3}\mathrm MeV~\mathrm fm^{-2}  &  \text{rank0} \\
	90A^{-5/3}\mathrm MeV~\mathrm fm^{-2}  &  \text{rank1} \\
	350A^{-5/3}\mathrm MeV~\mathrm fm^{-2} &  \text{rank2} \\
\end{cases}
$$
The values of the interaction constants were fixed from the experimental value of the resonance energies. The $A^{\dag}_{j_{p}j_{n}}(\lambda-\mu)$ and $A_{j_{p}j_{n}}(\lambda\mu)$ are the quasi-boson creation and annihilation operators, respectively, defined by
\begin{eqnarray}
	A^{\dag}_{j_{p}j_{n}}(\lambda-\mu)=
	\sqrt{\frac{2\lambda+1}{2j_{n}+1}}\sum_{m_{n}m_{p}}(-1)^{j_{p}-m_{p}}\alpha^{\dag}_{j_{p}m_{p}}\alpha_{j_{n}-m_{n}},
\end{eqnarray}
and
\begin{eqnarray}
	A_{j_{p}j_{n}}(\lambda\mu)=[A^{\dag}_{j_{p}j_{n}}(\lambda-\mu)]^{\dag}.\nonumber
\end{eqnarray}
The ${b}_{j_{p}j_{n}}(\lambda)$, $\bar{b}_{j_{p}j_{n}}(\lambda)$ in Eq.~(3) stand for the reduced matrix elements of the related multipole operators.

For $\Delta J^{\pi}= 1^{+}$, transitions $(\lambda=1)$ were given by
\begin{eqnarray}
	\bar b_{j_pj_n}(\lambda)= \frac{1}{\sqrt{2\lambda+1}}\langle j_p(l_ps_p)||\sigma_{k} \tau^{\pm}_{k} ||j_n(l_ns_n) \rangle \upsilon_{j_n}u_{j_p}.	\nonumber
\end{eqnarray}

For $\Delta J^{\pi}= 0^{-},1^{-},2^{-}$, transitions were described in a general form by 
\begin{eqnarray}
	\bar b_{j_pj_n}(\lambda)= \frac{1}{\sqrt{2\lambda+1}}\langle j_p(l_ps_p)|| \tau^{\pm}_{k}r_{k}[Y_{1}(\hat{r}_{k})\sigma_{k}]_{\lambda} ||j_n(l_ns_n) \rangle \upsilon_{j_n}u_{j_p}	\nonumber
\end{eqnarray}	
\begin{eqnarray}
	b_{j_pj_n}(\lambda)=\frac{\bar{b}_{j_pj_n}(\lambda)}{\upsilon_{j_n}u_{j_p}}u_{j_n}\upsilon_{j_p}, \nonumber
\end{eqnarray}
where $\upsilon$ and ${u}$ are single-particle and hole amplitudes, respectively.

The solution of  Hamiltonian Eq.~(1) is briefly described below. The allowed GT and the charge-exchange vibration modes in odd-odd nuclei are considered as the phonon excitations and described by
\begin{eqnarray}
	\mid\Psi_{i}>=Q^{\dag}_{i}\mid0>=\Sigma_{j_{\tau}\mu}[\psi^{i}_{j_{p}j_{n}}A^{\dag}(\lambda\mu)_{j_{p}j_{n}}-\varphi^{i}_{j_{p}j_{n}}A_{j_{p}j_{n}}(\lambda\mu)]\mid0>, \nonumber
\end{eqnarray}
where $Q^{\dag}_{i}$ is the pn-QRPA phonon creation operator, $\mid0>$ is the phonon vacuum which corresponds to the ground state of an even-even nucleus and fulfills $Q_{i}\mid0>$=0 for all \textit{i}. The $\psi^{i}_{j_{p}j_{n}}$ and $\varphi^{i}_{j_{p}j_{n}}$ are forward and backward quasiboson amplitudes. Employing the conventional procedure of the pn-QRPA we solved the equation of motion
\begin{eqnarray}
	[H, Q^{\dag}_{i}]\mid0>=\omega_{i}Q^{\dag}_{i}\mid0>. \nonumber
\end{eqnarray}
The $\omega_{i}$ is the $\textit{i}$th $0^{-}$, $1^{-}$, and $2^{-}$ excitation energy in odd-odd daughter calculated from the ground state of the parent even-even nucleus. Further details of the formalism can be seen in refs.~\cite{Nab16,Nab17,Cak10a,Cak12,Cak10,Nab15}. 

\subsubsection{Extension in the model for odd-A Nuclei}
We present a brief summary of the necessary formalism for treating odd-A nuclei in our model. The Hamiltonian of the odd-A system was chosen as	
\begin{eqnarray}
	H_{SSM}=H_{sqp}+h_{ph}^{CC}+h_{ph}^{CD}.
\end{eqnarray}
The effective residual interactions were determined using 
\begin{eqnarray}
	h_{ph}^{CC}=\frac{2\chi_{ph}}{g_{A}^2}\sum_{j_{\tau}j_{\tau'}\mu\mu'}[\bar b_{j_{p}j_{n}}(\lambda)C^{\dag}_{j_{p}j_{n}}(\lambda\mu)+(-1)^{(\lambda+\mu)}{b}_{j_{p}j_{n}}(\lambda)C_{j_{p}j_{n}}(\lambda-\mu)] \nonumber
\end{eqnarray}
\begin{eqnarray}
	\times[\bar b_{j_{p'}j_{n'}}(\lambda)C_{j_{p'}j_{n'}}(\lambda\mu')+{b}_{j_{p'}j_{n'}}(\lambda)C^{\dag}_{j_{p'}j_{n'}}(\lambda-\mu')]
\end{eqnarray}
\begin{eqnarray}
	h_{ph}^{CD}=\frac{2\chi_{ph}}{g_{A}^2}\sum_{j_{\tau}j_{\tau'}\mu\mu'}[\bar b_{j_{p}j_{n}}(\lambda)C^{\dag}_{j_{p}j_{n}}(\lambda\mu)+(-1)^{(\lambda+\mu)}{b}_{j_{p}j_{n}}(\lambda)C_{j_{p}j_{n}}(\lambda-\mu)] \nonumber
\end{eqnarray}
\begin{eqnarray}
	\times[d_{j_{p'}j_{n'}}(\lambda)D_{j_{p'}j_{n'}}(\lambda\mu')+\bar {d}_{j_{p'}j_{n'}}(\lambda)D^{\dag}_{j_{p'}j_{n'}}(\lambda-\mu')].
\end{eqnarray}
The $C^{\dag}_{j_{p}j_{n}}(\lambda-\mu)$, $D^{\dag}_{j_{p}j_{n}}(\lambda-\mu)$ and $C_{j_{p}j_{n}}(\lambda\mu)$, $D_{j_{p}j_{n}}(\lambda\mu)$ are the quasi-boson creation and annihilation operators, respectively, and given as
\begin{eqnarray}
	C^{\dag}_{j_{p}j_{n}}(\lambda-\mu)=
	\sqrt{\frac{2\lambda+1}{2j_{n}+1}}\sum_{m_{n}m_{p}}(-1)^{j_{p}-m_{p}}\alpha^{\dag}_{j_{n}m_{n}}\alpha^{\dag}_{j_{p}-m_{p}},\nonumber
\end{eqnarray}
and
\begin{eqnarray}
	C_{j_{p}j_{n}}(\lambda\mu)=[C^{\dag}_{j_{p}j_{n}}(\lambda-\mu)]^{\dag}\nonumber
\end{eqnarray}
\begin{eqnarray}
	D^{\dag}_{j_{p}j_{n}}(\lambda-\mu)=
	\sqrt{\frac{2\lambda+1}{2j_{n}+1}}\sum_{m_{n}m_{p}}(-1)^{j_{p}-m_{p}}\alpha^{\dag}_{j_{n}m_{n}}\alpha_{j_{p}-m_{p}},\nonumber
\end{eqnarray}
and
\begin{eqnarray}
	D_{j_{p}j_{n}}(\lambda\mu)=[D^{\dag}_{j_{p}j_{n}}(\lambda-\mu)]^{\dag}.\nonumber
\end{eqnarray}	
The ${b}_{j_{p}j_{n}}(\lambda)$, $\bar{b}_{j_{p}j_{n}}(\lambda)$, ${d}_{j_{p}j_{n}}(\lambda)$, $\bar{d}_{j_{p}j_{n}}(\lambda)$ in Eqs.~(6) and (7) stand for the reduced matrix elements of the related multipole operators. The charge-exchange spin-spin and spin-dipole transitions were defined by

\begin{eqnarray}
	\bar d_{j_pj_n}(\lambda)= \frac{1}{\sqrt{2\lambda+1}}\langle j_p(l_ps_p)||\sigma_{k} \tau^{\pm}_{k} ||j_n(l_ns_n) \rangle \upsilon_{j_n}\upsilon_{j_p}	\nonumber
\end{eqnarray}
\begin{eqnarray}
	\bar d_{j_pj_n}(\lambda)= \frac{1}{\sqrt{2\lambda+1}}\langle j_p(l_ps_p)|| \tau^{\pm}_{k}r_{k}[Y_{1}(\hat{r}_{k})\sigma_{k}]_{\lambda} ||j_n(l_ns_n) \rangle \upsilon_{j_n}\upsilon_{j_p},	\nonumber
\end{eqnarray}	
and
\begin{eqnarray}
	d_{j_pj_n}(\lambda)=\frac{\bar{d}_{j_pj_n}(\lambda)}{\upsilon_{j_n}\upsilon_{j_p}}u_{j_n}u_{j_p}. \nonumber
\end{eqnarray}
The wave function of the odd-A (with odd neutrons) nuclei is given by 
\begin{eqnarray}
	\arrowvert\Psi^{j}_{j_{k}m_{k}}>=\Omega^{j\dag}_{j_{k}m_{k}}\arrowvert 0>= 
	(N^{j}_{j_{k}}\alpha^{\dag}_{j_{k}m_{k}}+\sum_{j_{\nu}m_{\nu}}R^{ij}_{k\nu}A^{\dag}_{i}\alpha^{\dag}_{j_{\nu}m_{\nu}})\arrowvert 0>, \nonumber
\end{eqnarray}
where $\Omega^{j\dag}_{j_{k}m_{k}}$ and $\mid0>$ represent the phonon operator and  phonon vacuum, respectively. $N^{j}_{j_{k}}$ and $R^{ij}_{k\nu}$ are the quasiboson amplitudes corresponding to the states and  fulfill the normalization conditions. The wave function was formed by superposition of one- and three-quasiparticle (one-quasiparticle $+$ one phonon) states. The equation of motion was given by 
\begin{eqnarray}
	[H_{SSM},\Omega^{j\dag}_{j_{k}m_{k}}] \arrowvert 0>=\omega^{j}_{j_{k}m_{k}}\Omega^{j\dag}_{j_{k}m_{k}}\arrowvert 0>.\nonumber
\end{eqnarray}
The excitation energies $\omega^{j}_{j_{k}m_{k}}$ and the wave functions of the GT and FF excitations were obtained from the pn-QRPA(WS) equation of motion. The details of solution for the odd-A nuclei can be seen from ref.~\cite{Cak10b}.

\subsubsection{Nuclear Matrix Elements}
The FF transition consists of six nuclear matrix elements (NMEs)~\cite{BM69}. The NMEs include relativistic and  non-relativistic terms for the $\Delta J^{\pi}= 0^{-}, 1^{-}$ transitions. The relativistic NMEs were calculated directly without any assumptions. The contribution from the spin-orbit potential was included in the calculation of the relativistic matrix elements. The unique first forbidden (U1F) transitions, $\Delta J^{\pi}= 2^{-}$, do not contain any relativistic term. The matrix elements of single quasiparticle were calculated using the WS radial wave functions. The parameters were taken from ref.~\cite {Sol76}. The proton and neutron pairing gaps were determined using $\Delta_{p}=C_{p}/\sqrt{A}$ and $\Delta_{n}=C_{n}/\sqrt{A}$, respectively \cite{BM69}. The pairing strength parameters $(C_{p}$ and $C_{n})$ were fixed to reproduce the experimental pairing gaps \cite{Mol92}. The quenching of the effective value of $g_{A}$, in first-forbidden decay in the lead region, was studied by Bohr and Mottelson \cite{BM69} in the $\xi-$ approximation. The wave functions were assumed to be dominated by  single-particle configurations around $^{208}$Pb. Contributions of the virtual $0^{-}$ and $1^{-}$ intermediate states were attained within the $\xi-$ approximation. In the current investigation, we have used the same approach for the north-eastern region of the double magic $^{208}$Pb nucleus.

The transitions probabilities $\hat{B}(\lambda^{\pi}=0^{-},1^{-},2^{-})$ were specified as~\cite{BM69}
\begin{eqnarray}
	\hat{B}(\lambda^{\pi}=0^{-})=|<0^{-}_{i}\|M^{rank0}\|0^{+}>|^{2},
\end{eqnarray}
where
\begin{eqnarray}
	M^{rank0}=M(\rho_{A},\lambda=0)-i\frac{m_{e}c}{h}\xi \hat{M}(j_{A},\kappa=1,\lambda=0),
\end{eqnarray}
with
\begin{eqnarray}
	M(\rho_{A},\lambda=0)=\frac{g_{A}}{{(4\pi)}^{1/2}c}\Sigma_{k}\tau^{\pm}_{k}~(\sigma_{k}\cdot\vartheta_{k}) \nonumber
\end{eqnarray}
\begin{eqnarray}
	M(j_{A},\kappa=1,\lambda=0)=g_{A}\Sigma_{k}\tau^{\pm}_{k}r_{k}[Y_{1}(\hat{r}_{k})\sigma_{k})]_{0\mu} \nonumber
\end{eqnarray}
\begin{eqnarray}
	\hat{B}(\lambda^{\pi}=1^{-}) = |<1^{-}_{i}\|M^{rank1}\|0^{+}>|^{2},
\end{eqnarray}
where
\begin{eqnarray}
	M^{rank1}=M(j_{V},\kappa=0,\lambda=1,\mu) \pm i\frac{m_{e}c}{\sqrt{3}\hslash}\xi M(\rho_{V},\lambda=1,\mu)\nonumber
\end{eqnarray}
\begin{eqnarray}
	~~~~~~~~ \pm i\sqrt{\frac{2}{3}}\frac{m_{e}c}{\hslash}\xi M(j_{A},\kappa=1,\lambda=1,\mu), 
\end{eqnarray}
with
\begin{eqnarray}
	M(j_{V},\kappa=0,\lambda=1,\mu)= \frac{g_{V}}{c \sqrt{4\pi}}\Sigma_{k}\tau^{\pm}_{k}~(\vartheta_{k})_{1\mu} \nonumber
\end{eqnarray}
\begin{eqnarray}
	M(\rho_{V},\lambda=1,\mu)=g_{V}\Sigma_{k}\tau^{\pm}_{k}r_{k}Y_{1\mu}(\hat{r}_{k}) \nonumber
\end{eqnarray}
\begin{eqnarray}
	M(j_{A},\kappa=1,\lambda=1,\mu)= g_{A}\Sigma_{k}\tau^{\pm}_{k}r_{k}[Y_{1}(\hat{r}_{k})\sigma_{k}]_{1\mu} \nonumber
\end{eqnarray}
\begin{eqnarray}
	\hat{B}(\lambda^{\pi}=2^{-})=|<2^{-}_{i}\|M^{rank2}\|0^{+}>|^{2},
\end{eqnarray}
with
\begin{eqnarray}
	M^{rank2}=M(j_{A},\kappa=1,\lambda=2,\mu)= g_{A}\Sigma_{k}\tau^{\pm}_{k}r_{k}[Y_{1}(\hat{r}_{k})\sigma_{k}]_{2\mu} \nonumber,
\end{eqnarray}
where $\sigma_{k}$, ${\vartheta}_{k}$, $Y_{1}(\hat{r}_{k})$ and $\tau^{\pm}_{k}$ denote the Pauli matrices, velocity, spherical harmonics and isospin creation (annihilation) operators, respectively. The $\rho_{V}$ ($\rho_{A}$) and ${j}_{V}$ (${j}_{A}$) are the vector (axial vector) charge and current densities associated with a single nucleon, respectively. These variables are linear and depend only on the space point \textbf{r}. They are independent of the velocity. For further details we refer to Appendix 3D-Beta Interaction in ref. \cite{BM69}. In Eqs.~(10) and (12), the upper and lower signs refer to $\beta^-$ and $\beta^+$  decays, respectively

The multipole operators considered to calculate the reduced NMEs of the first forbidden transitions were defined using
\begin{eqnarray}
	M(j_{A},\kappa=1,\lambda,\mu)= g_{A}\Sigma_{k}\tau^{\pm}_{k} r_{k}[Y_{1}(\hat{r}_{k})\sigma_{k}]_{\lambda\mu},
\end{eqnarray}
where $\lambda=0, 1, 2$ and $\mu=(0,\pm1,\pm2,\dots, \pm\lambda)$ are the corresponding nuclear spin $(\lambda^{\pi}=0^{-},1^{-},2^{-})$ for the transition and its projection, respectively. The $ft$ values of the GT and non-unique first forbidden transitions were calculated using
\begin{eqnarray}
	ft=\frac{D}{(g_{A}/g_{V})^{2}4\pi~\hat{B}(\lambda^{\pi}=1^{+},0^{-},1^{-})}.
\end{eqnarray}
and for U1F transitions using \cite{BM69} 
\begin{eqnarray}
	ft=\frac{(2n+1)!!}{(n+1!)^2n!}~\frac{D}{(g_{A}/g_{V})^{2}4\pi~\hat{B}(\lambda^{\pi}=2^{-})},
\end{eqnarray}
where ${D}$ is a constant taken as 6295$s$. The effective ratio of axial and vector coupling constant was taken as $(g_{A}/g_{V})=-1.254$ \cite{War94}. We performed our schematic model calculation, based on the shell model, with and without a quenching factor. We took $(g_{A}/g_{V})^2_{eff}=0.7(g_{A}/g_{V})^2$ as our quenching factor also used in ref.~\cite{Civ86}. The same quenching factor was used to calculated allowed and FF rates.
	%%%%%%%%%%%%%%%%%%%%%%%%%%%%%%%
	\subsection{The  pn-QRPA(N) Model}
	$\beta$-decay rates of neutron-rich Bi and Pb isotopes can be of utility for investigation of the $r$-process. As stated earlier, $r$-process occurs mostly in explosive environment possessing  temperatures of the order of $GK$. In this subsection we introduce yet another QRPA model, which we call pn-QRPA(N), that is capable of microscopically calculating allowed and unique first forbidden (U1F) transitions in stellar matter.	Prevailing temperatures within the stellar core can reach few tens of $GK$ and under these extreme conditions there exists a finite occupation probability of parent excited states. The total stellar rates are then achieved by summing over all the partial rates from parent levels until satisfactory convergence is achieved in the rate calculation.  The stellar rates were calculated using a separable multi-shell schematic interaction on top of axially symmetric-deformed mean-field calculation. This pn-QRPA model deals with a simple pairing plus quadrupole Hamiltonian with the incorporation of particle-hole ($ph$) and particle-particle ($pp$) GT forces of separable form. These forces  were parameterized by interaction constants $\bar{\chi}$ and $\bar{\kappa}$, respectively.   The deformed Nilsson oscillator potential, with a quadratic deformation, was used to calculate the single-particle energies. Pairing gaps between like nucleons were chosen as given in the previous sub-section.  The necessary modification in the model to calculate allowed stellar weak rates  can be seen  from refs.~\cite{Mut92,Nab99,Nab04}. We briefly describe the formalism of U1F $\beta$ decay rates using the pn-QRPA(N) model below. 
	
	The stellar U1F weak rates from the $i^{th}$ parent state to the $f^{th}$ daughter state of the nucleus is given by
	\begin{center}
		\begin{eqnarray}
			\lambda_{if}=\frac{m^5_ec^4}{2\pi^3h^7}\sum_{\Delta J^\pi} g^2\phi(\Delta J^\pi;if) \hat{B}(\Delta J^\pi; if),
		\end{eqnarray}
	\end{center}
	where $\phi$($\Delta$J$^\pi$; $if$)  and $\hat{B}$($\Delta$J$^\pi$; $if$) represent integrated Fermi functions and  $\beta$-decay reduced transition probabilities, respectively, for the transition $i \rightarrow f$ that causes a change in spin and parity. The $g$ represents weak coupling constant. The reduce transition probabilities were calculated using 
	\begin{eqnarray}
		\hat{B}(\Delta J^\pi; if) = \frac{1}{12} z^2(\varOmega^2_m - 1) - \frac{1}{6} z^2\varOmega_m 
		\varOmega + \frac{1}{6} z^2\varOmega^2, 
	\end{eqnarray}
	where
	\begin{eqnarray}
		z = 2g_A \frac{<f||\sum_{k}r_k[\boldsymbol{C}^k_1 \times \boldsymbol{\sigma}]^2 \boldsymbol{\tau}^k_-|| i>}{\sqrt{2j_i + 1}},
	\end{eqnarray}
	and
	\begin{eqnarray}
		\boldsymbol{C}_{lm} = \sqrt{\frac{4\pi}{2l + 1}} \boldsymbol{Y}_{lm},
	\end{eqnarray}
	For the determination of the charge-changing U1F transitions, the  nuclear matrix elements were given by
	\begin{center}
		\begin{eqnarray}
			\Phi^{ph}_{\pi\nu,\pi^{'}\nu^{'}} = +2\bar{\chi}_{U1F} f_{\pi\nu}(\zeta)f_{\pi^{'}\nu^{'}}(\zeta)
		\end{eqnarray}
		\begin{eqnarray}
			\Phi^{pp}_{\pi\nu,\pi^{'}\nu^{'}} = -2\bar{\kappa}_{U1F} f_{\pi\nu}(\zeta)f_{\pi^{'}\nu^{'}}(\zeta),
		\end{eqnarray}
	\end{center}
	where
	\begin{center}
		\begin{eqnarray}
			f_{\pi\nu}(\zeta) = <j_{\pi}m_{\pi}|\tau\_r[\sigma Y_1]_{{2}{\zeta}}|j_{\nu}m_{\nu}>,
		\end{eqnarray}
	\end{center}
is a single-particle U1F transition amplitude. Here $\zeta$ denotes spherical component of the GT transition operator ($\zeta$ = 0, $\pm$1 and $\pm$2). It is to be noted that the neutron and proton states have opposite parities. The remaining symbols have their usual meanings. The interaction constants $\bar{\chi}_{U1F}$ and $\bar{\kappa}_{U1F}$ were chosen according to the recipe given in ref.~\cite{Hom96}. Deformation parameter of the selected nuclei was computed using the relation
	 \begin{center}
	 	\begin{eqnarray}
	 		\delta={125(Q_2)(Z)^{-1}(A)^{-2/3}}/{1.44},
	 	\end{eqnarray}
	 \end{center}
	 where $Z$ and $A$ are charge and mass numbers, respectively. The electric quadrupole moment ($Q_2$) values were taken from ref.~\cite{Mol16}. $Q$-values were computed using the recent mass compilation~\cite{Aud21}. 
	 
 We assumed that that the existing  temperature in the core of massive stars was high enough to  ionize the atoms completely. The electrons were not bound anymore to the nucleus and obeyed the Fermi–Dirac distribution. In our calculations, at high temperatures ($kT >$ 1 MeV), positrons were assumed to  appear via electron–positron pair creation. The positrons were further assumed to follow the same energy distribution function as the electrons. The integrated Fermi functions for stellar $\beta$-decay rates  were computed using
	\begin{eqnarray}
		\phi^{bd} =\int_{1}^{\varOmega_m} \varOmega \sqrt{\varOmega^2-1}(\varOmega_m-\varOmega)^2 [(\varOmega_m-\varOmega)^2 F_1(Z,\varOmega)+(\varOmega^2-1)F_2(Z,\varOmega)](1-P_-) d\varOmega,
	\end{eqnarray}
and
	\begin{eqnarray}
	\phi^{pc} =\int_{\varOmega_l}^{\infty} \varOmega \sqrt{\varOmega^2-1}(\varOmega_m+\varOmega)^2 [(\varOmega_m+\varOmega)^2 F_1(-Z,\varOmega)+(\varOmega^2-1)F_2(-Z,\varOmega)](1-P_+) d\varOmega,
\end{eqnarray}
for the positron capture rates.
In Eqs.~(24 - 25) we used natural units $(\hbar=c=m_{e}=1)$  where $m_{e}$ is the electron mass. $\varOmega$ is the total energy of the fermions (kinetic energy and rest mass). $\varOmega_l$ is the total capture threshold energy (rest + kinetic) for positron capture. If the corresponding electron emission total energy, $\varOmega_m$, is greater than -1, then
	$\varOmega_l$ = 1, and if it is less than or equal to 1, then $\varOmega_l = |\varOmega_m|$. $\varOmega$$_m$ is the total $\beta$-decay energy ($\varOmega_m$=$m_p$-$m_d$+$E_i$-$E_f$), where $m_p$ ($m_d$) and $E_i$ ($E_f$)  stand for mass and excitation energy of parent (daughter) nucleus.  The Fermi functions,
	$F_1(\pm Z,w)$ and $F_2(\pm Z,w)$, appearing in Eqs.~(24 - 25), were computed
	according to the procedure adopted by ref.~\cite{Gov71}.  The electron and positron distribution functions were determined using
	\begin{center}
		\begin{eqnarray}
			P_- = \frac{1}{[\exp(\frac{E-E_f}{kT})+1]}
		\end{eqnarray}
	\end{center}
	\begin{center}
	\begin{eqnarray}
		P_+ = \frac{1}{[\exp(\frac{E+2+E_f}{kT})+1]},
	\end{eqnarray}
\end{center}
where E = ($\varOmega - 1$) and $E_f$ are the kinetic and Fermi energies of electrons (including the rest mass), respectively. $T$ is the core temperature and $k$ is the Boltzmann constant. The electron number density related to nuclei and protons was determined using
	\begin{center}
		\begin{eqnarray}
			\rho Y_e = \frac{1}{\pi^2 N_A} (\frac{m_ec}{h})^3 \int_{0}^{\infty}(P_--P_+)p^2 dp,
		\end{eqnarray}
	\end{center}
where $\rho$ stands for baryon density, $Y_e$ represents the ratio of electron number to the baryon number, $N_A$ is the Avogadro's number,  $p$ = ($\varOmega$$^2$-1)$^{1/2}$ denotes the positron/electron momentum. Eq.~(28) was used to calculate the Fermi energies at specified temperature and $\rho$Y$_e$ values. 
The occupation probability of the parent states were calculated assuming the Boltzmann distribution. The rate per unit time per nucleus, for $\beta$-decay and positron capture, was finally calculated using
	\begin{center}
		\begin{eqnarray}
			\lambda = \sum_{if}P_i\lambda_{if}
		\end{eqnarray}
	\end{center}
The summation in Eq.~(29) was carried until a satisfactory convergence in the rate calculations was achieved. A large model space (up to 7 major shells) assisted to achieve the desired convergence in our rate calculations. No explicit quenching factor was introduced, following the recipe used in earlier stellar rate calculations~\cite{Nab99,Nab04}
	
\section{Results and Discussion}
\label{sec:3}
Calculations of $\beta$-decay half-lives normally introduce a quenching factor to reproduce measured data. Two important factors to justify quenching of the calculated GT strength are coupling of the weak  forces to two nucleons and  strong correlations within the nucleus~\cite{Gys19}. As stated earlier, we performed our terrestrial calculation both with and without an explicit quenching factor.

The calculated and measured log$ft$ values for the state-by-state transitions are shown in Tables (\ref{tab1}~-~\ref{tab2}). The results for $^{210-215}$Pb \textrightarrow $^{210-215}$Bi transitions are displayed in Table \ref{tab1} whereas Table \ref{tab2}  presents the results for $^{210-215}$Bi \textrightarrow $^{210-215}$Po transitions. Measured data was taken from ref.~\cite{Nudat} and shown in Column~5 of the tables. Columns (6~-~7) show our model results using the  pn-QRPA(WS) model (without quenching factor) and pn-QRPA$(WS)^*$ (with quenching factor). The last two columns show the shell model results~\cite{Sha22}. The authors used the KHPE interaction and performed the calculations without truncating the model space. The SM I calculation used a quenching factor of 0.38 for $g_A$ whereas SM II used quenching factors of 0.38 and 0.51 for $g_A$ and $g_V$, respectively. It is noted that introducing quenching values had a more pronounced effect in shell model calculations when compared with our results. The dashes in the table show the inability of the QRPA model to calculate the specified transitions.  It is noted from Tables (\ref{tab1}~-~\ref{tab2}) that the pn-QRPA(WS) model results compare well with the measured log$ft$ values. A large model space contributed to a reliable calculation of nuclear matrix elements of these heavy nuclei.  The restrictions on the matrix elements of \textbf{r} and its derivatives limit the first-forbidden decay to transitions between adjoint major shells. The calculated pn-QRPA (WS) log$ft$ values for the FF transitions depend whether the transition is between one or two quasiparticle states. This is because a single-phonon state wave function is a superposition of two quasiparticle wave functions, of which only one has a large weight coefficient. The microscopic structure of the charge-exchange spin-dipole transition strength distribution spectra of the calculated  pn-QRPA(WS) results (without quenching factor) for selected Pb \textrightarrow Bi and Bi \textrightarrow Po transitions are given in Tables (\ref{tab3}~-~\ref{tab4}). The last column in these tables shows the $\psi_{np}^{i}$ amplitude of collective states.  The energy dependence of $\beta$-decay probabilities of the FF  transitions are generally between (3 - 26)~$MeV$. The dominant contributions for the collective $0^{-}$, $1^{-}$, and $2^{-}$ states, in the north-east region of Pb-208 nuclei, are located at energies of the order (18 - 25)~$MeV$, (19 - 26)~$MeV$, and (19 - 24)~$MeV$, respectively. The effect of the giant-resonance is taken into consideration by using the sum rules. In our calculation the sum rule is fulfilled to around $87\%$.

Comparison of the model calculation with measured half-lives~\cite{Nudat} is shown in Table~\ref{tab5} and Table~\ref{tab6} for Pb and Bi isotopes, respectively.  It is clearly seen from the tables that our model results are in decent comparison with the measured data. Our model produced better results than the shell model calculations. In order to quantify performance of the models, we introduce the variable $X$ which gives the ratio of the measured  to calculated half-lives. $X$ is defined as 
\begin{equation}\label{eq54}
	X = \begin{array}{rcl}
		T^{cal}_{1/2}/T^{exp}_{1/2} 	~~~~~if~~~~~ T^{cal}_{1/2}\geq T^{exp}_{1/2} \\
		T^{exp}_{1/2}/T^{cal}_{1/2} 	~~~~~if~~~~~ T^{exp}_{1/2} > T^{cal}_{1/2}
	\end{array} .
\end{equation}
We next define the mean deviation, $\bar{X}$  
\begin{equation}\label{eq53}
	~ ~ ~ ~ ~ ~ ~ ~ ~ ~ ~ ~ ~ ~ ~ ~ ~ ~ ~ ~	\bar{X} = X/n,
\end{equation}
where $n$ is the number of nuclei ($n$ =  6 in the current investigation).
Table \ref{tab7} shows the calculated ratios. Here $X_{WS}$  ($X_{(WS)}$*) denotes the ratios using the pn-QRPA(WS) model without (with) quenching factor. $X_{SM I}$ and $X_{SM II}$ show the computed ratios using the SM I and SM II models, respectively. Except for $^{215}$Pb, all computed ratios using the pn-QRPA(WS) model are less than 2 which we consider a big triumph of the model. At the bottom of Table \ref{tab7}, we show the values of computed mean deviation using the different models. The $\bar{X}$ of pn-QRPA(WS) model (with and without quenching) is less than 2. This shows that the pn-QRPA(WS) model has a strong predictive power for calculation of $\beta$-decay half-lives for heavy neutron-rich nuclei. The computed mean deviations validate our previous comment that quenching factors affected the shell model results more than the pn-QRPA(WS) model.

Realistic estimate of heavy element abundance distributions are a pre-requisite for use as initial conditions in explosive $r$-process calculations~\cite{Bur57}. We present, for the first time, weak rates  for Pb and Bi isotopes in stellar environment. These include $\beta$-decay and positron capture rates. We present both allowed GT and U1F rates, for neutron-rich Pb and Bi nuclei, using the pn-QRPA(N) model. It is reminded that the pn-QRPA(N) calculated stellar rates may only be taken as a rough estimate as these do not include rank1 and rank2 FF transitions. Incorporation of FF rates in the model would be taken as a future assignment. The rates are shown in Tables (\ref{tab8}~-~\ref{tab13}) at selected values of core density and temperature.  The choices of densities agree approximately with the regime just outside of solar hydrogen burning, with helium burning and with the conditions of core collapse prior to neutrino trapping. It is noted from the tables that the rates increase as the core temperature rises and decrease as density of the stellar core increases by orders of magnitude. The available phase space for the reactions decreases with increasing stellar densities. On the other hand, increasing the core temperature tends to weaken the Pauli blocking and enhances the rates because of finite contribution from parent excited states. As the temperature rises, the contribution to the total rate from positron capture increases as more positrons are created at high temperatures ($kT >$ 1 MeV). The U1F phase space grows significantly as temperature of the core increases and has a sizeable contribution to the total phase space. For Bi and Pb isotopes, the U1F phase space increases appreciably with increasing neutron number.  It is further noted from the tables that increasing the core temperature affects the allowed GT rates more than the U1F rates. This effect may be explained when one recalls that the first-forbidden transitions occur at relatively high energies and are primarily of particle-hole nature. Smearing of single-particle strength functions due to increasing temperatures and pairing correlations affect them relatively less. The emitted electron in $\beta$-decay can achieve any kinetic energy  ranging from zero up to the maximum  available for the decay. In extreme stellar environment, possessing high electron density, fewer states are admissible for emission of low-energy electrons. In other words, allowed $\beta$-decay is strongly suppressed if maximum kinetic energy for the decay is small (as in the current investigation). On the other hand, for U1F decays, the inhibition of decays is not as big as that of allowed states for the same density, temperature and $\beta$-decay energy~\cite{Bor05}. Our results show that for many nuclei there exits a 100\% contribution from the U1F rates. Our findings are in line with the earlier reported calculations in refs.~\cite{Hom96,Nab17}.

\section{Conclusion}
\label{sec:4} The ISOLDE Decay Station at the CERN-ISOLDE facility and new generation radioactive ion-beam facilities (e.g. RIBF (Japan), FRIB (USA) and FAIR(Germany)) will lead to identification of new $\beta$-decaying states and will make accessible measured GT strength functions of many more exotic (including neutron-rich) nuclei. This in turn will pose constraints and new challenges for theoretical models. The region around doubly magic nucleus (e.g., $^{208}$Pb), due to its increased stability, has always attracted attention for further investigations. The $\beta$-decay in the lead region demands theoretical predictions of $\beta$-decay observables, as the measurement of half-lives of nuclei near N = 126 is a formidable task. It is well established that the $r$-process path lies far from accessible nuclei. The FF transitions contribute well to the total $\beta$ decay half-lives for neutron-rich isotopes. We used the spherical pn-QRPA(WS) to calculate allowed GT and FF transitions. The deformed pn-QRPA(N) was used to examine the allowed GT $\&$ U1F  stellar rates pertinent to $r$-process conditions.  We performed the calculations for neutron-rich Bi and Pb isotopes for the mass range A = (210~-~215). 

The log $ft$ values and half-lives obtained by the pn-QRPA(WS) calculation showed a decent agreement with the available experimental results. The set of log$ft$ values obtained using the pn-QRPA$(WS)^*$ model, adopting a quenching factor of 0.7, gave the best results. Our investigation of stellar rates revealed that increasing core temperature affected the GT rates more than the U1F rates. It was concluded that, for certain nuclei, the U1F contribution is significantly more than the allowed rates under stellar conditions.  

The calculations support the argument that the pn-QRPA model gives reliable prediction for neutron-rich nuclei \cite{Sta90,Hom96}. The reported $\beta$-decay observable bear consequences for the site-independent $r$-process calculation and the nucleosynthesis problem. The calculations might contribute in accelerating the $r$-matter flow relative to calculations based only on allowed GT rates. Core-collapse simulators are encouraged to test-run the reported rates for possible interesting outcomes in their results.

	%%%%%%%%%%%%%%%%%%%%%% Acknowledgment %%%%%%%%%%%%%%%%%%%%%%%%%%

\vspace{0.5in} \textbf{Acknowledgment}:  J.-U. Nabi and A. Mehmood would like to acknowledge the support of the Higher Education Commission Pakistan through project number 20-15394/NRPU/R$\&$D/HEC/2021.

	%%%%%%%%%%%%%%%%%%%%%%%%%%% References %%%%%%%%%%%%%%%%%%%%%%%%%%%%%
	
	%%%%%%%%%%%%%%%%%%%%TABLES%%%%%%%%%%%%%%%%%%%%%
	\clearpage
	\begin{table}
	\centering
	\caption{Comparison between theoretical and experimental log $ft$ values for Pb \textrightarrow Bi transitions. The pn-QRPA(WS)$^{*}$ (pn-QRPA(WS)) calculation is with (without) a quenching factor. Measured data  was taken from ref. \cite{Nudat}. Last two columns show  shell model data taken from ref. \cite{Sha22}. FNU and FU stand for Forbidden Non-Unique and Forbidden Unique transitions, respectively.}\label{tab1}
		\scriptsize
\renewcommand{\arraystretch}{2.0}
\renewcommand{\tabcolsep}{0.12cm}
\renewcommand{\ULdepth}{6.0pt}
\vspace{0.25cm}
\begin{tabular}{cccccccccc}
	\hline
	\multicolumn{2}{c}{\uline{~~~~~~~~Transition~~~~~~~~}} &
	\multicolumn{1}{c}{Decay Mode} &
	\multicolumn{1}{c}{E~(keV)}     &
	\multicolumn{6}{c}{\uline{~~~~~~~~~~~~~~~~~~~~~~~~~~~~~~~~~~~~~ log $ft$ ~~~~~~~~~~~~~~~~~~~~~~~~~~~~~~}} \\
	
	\multicolumn{1}{c}{Initial ($J_i^{\pi}$) } &
	\multicolumn{1}{c}{Final ($J_f^{\pi}$)}  &
	\multicolumn{1}{c}{~~  ~~} &
	\multicolumn{1}{c}{~~  ~~} &
	\multicolumn{1}{c}{~Exp.} &
	\multicolumn{1}{c}{~pn-QRPA(WS)} &
	\multicolumn{1}{c}{~pn-QRPA(WS)$^{*}$} &
	\multicolumn{1}{c}{SM I} &
	\multicolumn{1}{c}{SM II} \\
	\hline
	
	$^{210}$Pb$(0^+)$ & $^{210}$Bi$(1_1^-)$ & 1st FNU & 0.0 & 7.9 & 7.06 & 7.36 & 6.719 & 8.517 \\
	
	& $^{210}$Bi$(0_1^-)$ & 1st FNU & 46.539 & 5.4 & 5.25 & 5.55 & 5.469 & 5.469 \\
	
	$^{211}$Pb$(9/2^+)$ & $^{211}$Bi$(9/2_1^-)$ & 1st FNU & 0.0 & 5.99 & 5.43 & 5.73 & 6.109 & 6.109 \\
	
	& $^{211}$Bi$(7/2_1^-)$ & 1st FNU & 404.866 & 7.19 & 6.81 & 7.11 & 6.239 & 7.183 \\
	
	& $^{211}$Bi$(9/2_2^-)$ & 1st FNU & 831.960 & 5.733 & 5.346 & 5.647 & 5.779 & 5.771 \\
	
	& $^{211}$Bi$(9/2_3^-)$ & 1st FNU & 1109.485 & 5.58 & 5.321 & 5.622  & 5.507 & 5.513 \\
	
	$^{212}$Pb$(0^+)$ & $^{212}$Bi$(1_1^-)$ & 1st FNU & 0.0 & 6.73 & 6.177 & 6.478 & 7.486 & 7.288 \\
	
	& $^{212}$Bi$(2_1^-)$ & 1st FU & 115.183 & $-$ & 9.251 & 9.502 & 10.143 & 10.140\\
	
	& $^{212}$Bi$(0_1^-)$ & 1st FNU & 238.632 & 5.179 & 5.137 & 5.438 & 5.208 & 5.208 \\
	
	& $^{212}$Bi$(1_2^-)$ & 1st FNU & 415.272 & 5.342 & 5.202 & 5.503 & 4.551 & 5.156 \\
	
	$^{213}$Pb$(9/2^+)$ & $^{213}$Bi$(9/2_1^-)$ & 1st FNU & 0.0 & 6.5 & 6.217 & 6.518 & 6.447 & 6.459 \\
	
	& $^{213}$Bi$(7/2_1^-)$ & 1st FNU & 257.63 & 7.7 & 6.582 & 6.883 & 6.719 & 7.887 \\
	
	& $^{213}$Bi$(5/2_1^-)$ & 1st FU & 592.72 & 7.5 & 7.728 & 8.128 & 13.355 & 13.698 \\
	
	& $^{213}$Bi$(7/2_2^-)$ & 1st FNU & 592.72 & 7.5 & 7.801 & 8.103 & 8.184 & 8.018 \\
	
	& $^{213}$Bi$(9/2_2^-)$ & 1st FNU & 977.71 & 5.6 & 5.871 & 6.172 & 6.128 & 6.107 \\
	
	$^{214}$Pb$(0^+)$ & $^{214}$Bi$(1_1^-)$ & 1st FNU & 0.0 & 6.26 & 6.419 & 6.720 & 7.873 & 7.144 \\
	
	& $^{214}$Bi$(2_1^-)$ & 1st FU & 53.2260 & ${-}$ & 8.213 & 8.514 & 9.446 & 9.445 \\
	
	& $^{214}$Bi$(2_2^-)$ & 1st FU & 62.68 & ${-}$ & 7.436 & 7.747 & 8.087 & 8.098 \\
	
	& $^{214}$Bi$(3_1^-)$ & 1st FNU & 62.68 & ${-}$ & ${-}$ & ${-}$ & 15.860 & 14.949 \\
	
	& $^{214}$Bi$(2_3^-)$ & 1st FU & 258.869 & 8.04 & 8.246 & 8.547 & 8.665 & 8.668 \\
	
	& $^{214}$Bi$(1_2^-)$ & 1st FNU & 295.2236 & 5.25 & 5.431 & 5.732 & 4.494 & 5.014 \\
	
	& $^{214}$Bi$(0_1^-)$ & 1st FNU & 351.9323 & 5.07 & 5.101 & 5.414 & 5.080 & 5.080 \\
	
	& $^{214}$Bi$(1_3^-)$ & 1st FNU & 351.9323 & 5.07 & 5.213 & 5.724 & 5.074 & 6.187 \\
	
	& $^{214}$Bi$(2_4^-)$ & 1st FU & 377.03 & ${-}$ & 9.354 & 9.665 & 10.211 & 10.197 \\
	
	& $^{214}$Bi$(1_4^-)$ & 1st FNU & 533.672 & 6.23 & 6.743 & 7.054 & 7.374 & 7.881 \\
	
	& $^{214}$Bi$(1_1^+)$ & Allowed & 838.994 & 4.43 & 4.357 & 4.668 & 4.238 & 4.238 \\
	
	$^{215}$Pb$(9/2^+)$ & $^{215}$Bi$(9/2_1^-)$ & 1st FNU & 0.0 & $\geq$ 6.1 & 6.532 & 6.833 & 6.713 & 6.741 \\
	
	& $^{215}$Bi$(7/2_1^-)$ & 1st FNU & 183.5 & $>$ 6.6 & 6.845 & 7.146 & 7.224 & 8.486 \\
	
	\hline
	
\end{tabular}
\end{table}
\clearpage

	\begin{table}
	\centering
	\caption{Same as Table~\ref{tab1} but for Bi \textrightarrow Po transitions.}\label{tab2}
		
		\scriptsize
\renewcommand{\arraystretch}{2.0}
\renewcommand{\tabcolsep}{0.12cm}
\renewcommand{\ULdepth}{6.0pt}
\vspace{0.25cm}
\begin{tabular}{cccccccccc}
	\hline
	\multicolumn{2}{c}{\uline{~~~~~~~~Transition~~~~~~~~}} &
	\multicolumn{1}{c}{Decay Mode} &
	\multicolumn{1}{c}{E~(keV)}     &
	\multicolumn{6}{c}{\uline{~~~~~~~~~~~~~~~~~~~~~~~~~~~~~~~~~~~~~ log $ft$ ~~~~~~~~~~~~~~~~~~~~~~~~~~~~~~}} \\
	\multicolumn{1}{c}{Initial ($J_i^{\pi}$) } &
	\multicolumn{1}{c}{Final ($J_f^{\pi}$)}  &
	\multicolumn{1}{c}{~~  ~~} &
	\multicolumn{1}{c}{~~  ~~} &
	\multicolumn{1}{c}{~Exp.} &
	\multicolumn{1}{c}{~pn-QRPA(WS)} &
	\multicolumn{1}{c}{~pn-QRPA(WS)$^{*}$} &
	\multicolumn{1}{c}{SM I} &
	\multicolumn{1}{c}{SM II} \\
	
	\hline
	
	$^{210}$Bi$(1^-)$ & $^{210}$Po$(0_1^+)$ & 1st FNU & 0.0 & 8.0 & 7.512 & 7.813 & 6.690 & 8.114 \\
	
	$^{211}$Bi$(9/2^-)$ & $^{211}$Po$(9/2_1^+)$ & 1st FNU & 0.0 & 5.99 & 5.81  & 6.111 & 6.143 & 6.140 \\
	
	$^{212}$Bi$(1^-)$ & $^{212}$Po$(0_1^+)$ & 1st FNU & 0.0 & 7.266 & 6.673 & 6.974 & 7.33 & 8.951 \\
	
	& $^{212}$Po$(2_1^+)$ & 1st FNU & 727.330 & 7.720 & $-$ & $-$ & 7.017 & 7.609 \\
	
	& $^{212}$Po$(2_2^+)$ & 1st FNU & 1512.70 & 7.093 & $-$ & $-$ & 6.554 & 6.645 \\
	
	& $^{212}$Po$(1_1^+)$ & 1st FNU & 1620.738 & 6.748 & $-$ & $-$ & 6.079 & 6.384 \\
	
	& $^{212}$Po$(2_3^+)$ & 1st FNU & 1679.450 & 7.51 & $-$ & $-$ & 5.85 & 6.447 \\
	
	& $^{212}$Po$(0_2^+)$ & 1st FNU & 1800.9 & 8.05 & 7.645 & 7.946 & 8.103 & 7.780 \\
	
	& $^{212}$Po$(2_4^+)$ & 1st FNU & 1805.96 & 6.695 & $-$ & $-$ & 6.546 & 6.926 \\
	
	$^{213}$Bi$(9/2^-)$ & $^{213}$Po$(9/2_1^+)$ & 1st FNU & 0.0 & 6.31 & 6.023 & 6.324 & 6.367 & 6.358 \\
	
	& $^{213}$Po$(11/2_1^+)$ & 1st FNU & 292.805 & 8.45 & 7.401 & 7.702 & 5.386 & 5.869 \\
	
	& $^{213}$Po$(7/2_1^+)$ & 1st FNU & 440.446 & 6.08 & 5.676 & 5.977 & 9.381 & 9.302 \\
	
	& $^{213}$Po$(5/2_1^+)$ & 1st FU & 600.87 & 10.03 & 9.345 & 9.646 & 13.513 & 13.462 \\
	
	& $^{213}$Po$(13/2_1^+)$ & 1st FU & 867.98 & 8.64 & 8.242 & 8.443 & 10.457 & 10.458 \\
	
	& $^{213}$Po$(9/2_2^+)$ & 1st FNU & 1003.605 & 7.49 & 7.152 & 7.453 & 6.096 & 6.091 \\
	
	& $^{213}$Po$(9/2_3^+)$ & 1st FNU & 1045.65 & 7.85 & 7.085 & 7.386 & 7.196 & 7.211 \\
	
	& $^{213}$Po$(11/2_2^+)$ & 1st FNU & 1045.65 & 7.85 & 7.324 & 7.625 & 8.293 & 8.591 \\
	
	$^{214}$Bi$(1^-)$ & $^{214}$Po$(0_1^+)$ & 1st FNU & 0.0 & 7.872 & 8.216 & 8.517 & 7.314 & 8.311 \\
	
	& $^{214}$Po$(2_1^+)$ & 1st FNU & 609.318 & 9.06 & $-$ & $-$ & 7.649 & 8.318 \\
	
	& $^{214}$Po$(3_1^+)$ & 2nd FNU & 1274.765 & 9.5 & $-$ & $-$ & 11.161 & 11.542 \\
	
	& $^{214}$Po$(2_2^+)$ & 1st FNU & 1377.681 & 7.374 & $-$ & $-$ & 6.695 & 6.978 \\
	
	& $^{214}$Po$(0_2^+)$ & 1st FNU & 1415.498 & 8.25 & 8.425 & 8.726 & 7.38 & 7.644 \\
	
	& $^{214}$Po$(2_3^+)$ & 1st FNU & 1661.282 & 8.21 & $-$ & $-$ & 6.302 & 6.929 \\
	
	& $^{214}$Po$(2_4^+)$ & 1st FNU & 1729.613 & 6.654 & $-$ & $-$ & 6.713 & 6.240 \\
	
	& $^{214}$Po$(2_5^+)$ & 1st FNU & 1847.446 & 6.859 & $-$ & $-$ & 6.509 & 7.179 \\
	
	& $^{214}$Po$(2_6^+)$ & 1st FNU & 2010.831 & 7.422 & $-$ & $-$ & 6.381 & 6.730 \\
	
	$^{215}$Bi$(9/2^-)$ & $^{215}$Po$(9/2_1^+)$ & 1st FNU & 0.0 & $>$ 6.9 & 7.201 & 7.503 & 6.653 & 6.642 \\
	
	& $^{215}$Po$(7/2_1^+)$ & 1st FNU & 271.11 & $>$ 8.2 & 9.376 & 9.677 & 10.020 & 8.737 \\
	
	& $^{215}$Po$(11/2_1^+)$ & 1st FNU & 293.53 & 6.0 & 6.245 & 6.546 & 5.349 & 5.813 \\
	
	& $^{215}$Po$(5/2_1^+)$ & 1st FU & 401.6 & 7.7 & 8.253 & 8.554 & 13.796 & 13.714 \\
	
	& $^{215}$Po$(7/2_2^+)$ & 1st FNU & 517.53 & 7.8 & 8.173 & 8.474 & 9.041 & 8.965 \\
	
	& $^{215}$Po$(9/2_2^+)$ & 1st FNU & 517.53 & 7.8 & 7.032 & 7.333 & 6.619 & 6.608 \\
	
	& $^{215}$Po$(11/2_2^+)$ & 1st FNU & 609.0 & 7.4 & 8.213 & 8.514 & 9.058 & 10.09 \\
	
	& $^{215}$Po$(13/2_1^+)$ & 1st FU & 609.0 & 7.4 & 7.912 & 8.211 & 9.931 & 9.933 \\
	
	\hline
	
\end{tabular}
\end{table}

	\begin{table}
	\centering
	\caption{The microscopic structure of the FF transition strength distributions of the pn-QRPA(WS) results (without quenching factor) for selected Pb \textrightarrow Bi transitions. $\omega_{i}$ denotes the excitation energy in daughter nucleus. The last column gives the wavefunction values of the parent states.}\label{tab3}
	\scriptsize
	\renewcommand{\arraystretch}{2.0}
	\renewcommand{\tabcolsep}{0.12cm}
	\renewcommand{\ULdepth}{6.0pt}
	\vspace{0.25cm}
	\begin{tabular}{ccccc}
		\hline			
		\multicolumn{1}{c}{Initial ($J_i^{\pi}$) } &
		\multicolumn{1}{c}{Final ($J_f^{\pi}$)}  &
		\multicolumn{1}{c}{~$\omega_{i}$ ($\mathrm MeV$)} &
		\multicolumn{1}{c}{~Structure} &
		\multicolumn{1}{c}{~$\psi_{np}^{i}$} \\
		\hline
		
		$^{210}$Pb$(0^+)$ & $^{210}$Bi$(0_{1}^-)$ & 18.17 & ($3p^{n}_{{1/2}}$~-~$4s^{p}_{{1/2}}$) & $-$0.96   \\
		&  & 18.57 & ($2d^{n}_{{3/2}}$~-~$3p^{p}_{{3/2}}$) &  $-$0.99 \\
		&  & 20.83 & ($1g^{n}_{{9/2}}$~-~$1h^{p}_{{9/2}}$) &  0.87  \\
		&  &  & ($2d^{n}_{{5/2}}$~-~$2f^{p}_{{5/2}}$) &  0.39  \\
		&  & 21.89 & ($2f^{n}_{{7/2}}$~-~$2g^{p}_{{7/2}}$) &  0.95  \\
		&  & 23.76 & ($1h^{n}_{{11/2}}$~-~$1i^{p}_{{11/2}}$) &  0.98 \\
		
		$^{210}$Pb$(0^+)$ & $^{210}$Bi$(1_{1}^-)$ & 19.07 & ($2d^{n}_{{3/2}}$~-~$3p^{p}_{{3/2}}$) &  $-$0.99 \\
		&  & 20.73 & ($2f^{n}_{{7/2}}$~-~$2g^{p}_{{7/2}}$) & 0.94  \\
		&  & 23.12 & ($1g^{n}_{{9/2}}$~-~$1h^{p}_{{9/2}}$) & 0.86  \\
		&  &       & ($3p^{n}_{{3/2}}$~-~$3d^{p}_{{3/2}}$) &  0.33 \\
		&  & 25.68 & ($1h^{n}_{{11/2}}$~-~$1i^{p}_{{11/2}}$) & 0.96 \\	    
		
		$^{212}$Pb$(0^+)$ & $^{212}$Bi$(0_{1}^-)$ & 18.35 & ($2f^{n}_{{5/2}}$~-~$3d^{p}_{{5/2}}$) &  $-$0.99 \\
		&  & 19.52 & ($3p^{n}_{{1/2}}$~-~$4s^{p}_{{1/2}}$) & $-$0.97 \\
		&  & 19.83 & ($2d^{n}_{{3/2}}$~-~$3p^{p}_{{3/2}}$) &  $-$0.99 \\
		&  & 21.33 & ($1g^{n}_{{9/2}}$~-~$1h^{p}_{{9/2}}$) &  0.78 \\
		&  & 22.74 & ($2f^{n}_{{7/2}}$~-~$2g^{p}_{{7/2}}$) &  0.96 \\
		&  & 23.53 & ($1h^{n}_{{11/2}}$~-~$1i^{p}_{{11/2}}$) & 0.93 \\
		&  & 23.87 & ($2g^{n}_{{7/2}}$~-~$3f^{p}_{{7/2}}$) &  $-$0.99 \\
		
		$^{212}$Pb$(0^+)$ & $^{212}$Bi$(1_{1}^-)$ & 19.67 & ($3p^{n}_{{1/2}}$~-~$4s^{p}_{{1/2}}$)  &  $-$0.98 \\
		&  & 20.73 & ($2d^{n}_{{3/2}}$~-~$3p^{p}_{{3/2}}$)  & $-$0.98  \\
		&  & 21.17 & ($1g^{n}_{{9/2}}$~-~$1h^{p}_{{9/2}}$)  &  0.87  \\
		&  & 22.38 & ($2f^{n}_{{7/2}}$~-~$2g^{p}_{{7/2}}$)  &  0.95 \\
		&  & 23.41 & ($1h^{n}_{{11/2}}$~-~$1i^{p}_{{11/2}}$)   & 0.96  \\
		&  & 24.52 & ($1h^{n}_{{11/2}}$~-~$1i^{p}_{{11/2}}$)   &  0.43  \\
		&  &       & ($2g^{n}_{{7/2}}$~-~$3f^{p}_{{7/2}}$)   & $-$0.98  \\
		&  &       & ($1g^{n}_{{9/2}}$~-~$1h^{p}_{{9/2}}$)    & 0.11  \\

		$^{212}$Pb$(0^+)$ & $^{212}$Bi$(2_{1}^-)$ & 21.10 & ($2d^{n}_{{3/2}}$~-~$3p^{p}_{{3/2}}$) & $-$0.99 \\
		&  & 22.13 & ($1g^{n}_{{9/2}}$~-~$1h^{p}_{{9/2}}$) & 0.82 \\
		&  & 22.75 & ($2f^{n}_{{7/2}}$~-~$2g^{p}_{{7/2}}$) & 0.97 \\
		&  & 23.58 & ($1h^{n}_{{11/2}}$~-~$1i^{p}_{{11/2}}$) & 0.95 \\
		&  & 23.91 & ($2g^{n}_{{7/2}}$~-~$3f^{p}_{{7/2}}$) &  $-$0.99 \\		
			\hline
	\end{tabular}
	\end{table}
	\newpage

	\begin{table}
	\centering
	Table 3 Continued:\\
	\scriptsize
	\renewcommand{\arraystretch}{2.0}
	\renewcommand{\tabcolsep}{0.12cm}
	\renewcommand{\ULdepth}{6.0pt}
	\vspace{0.25cm}
	\begin{tabular}{ccccc}
		\hline			
		\multicolumn{1}{c}{Initial ($J_i^{\pi}$) } &
		\multicolumn{1}{c}{Final ($J_f^{\pi}$)}  &
		\multicolumn{1}{c}{~$\omega_{i}$ ($\mathrm MeV$)} &
		\multicolumn{1}{c}{~Structure} &
		\multicolumn{1}{c}{~$\psi_{np}^{i}$} \\
		\hline
		
		$^{214}$Pb$(0^+)$ & $^{214}$Bi$(0_{1}^-)$ & 18.54 & ($2f^{n}_{{5/2}}$~-~$3d^{p}_{{5/2}}$) & $-$0.99   \\
		&  & 18.92 & ($2d^{n}_{{3/2}}$~-~$3p^{p}_{{3/2}}$) & $-$0.99   \\
		&  & 19.37 & ($3p^{n}_{{3/2}}$~-~$3d^{p}_{{3/2}}$) &  0.97   \\
		&  & 20.31 & ($1g^{n}_{{9/2}}$~-~$1h^{p}_{{9/2}}$) &  0.86   \\
		&  & 22.78 & ($2f^{n}_{{7/2}}$~-~$2g^{p}_{{7/2}}$) &  0.95   \\
		&  & 23.88 & ($1h^{n}_{{11/2}}$~-~$1i^{p}_{{11/2}}$) &  0.98   \\			
		
		$^{214}$Pb$(0^+)$ & $^{214}$Bi$(1_{1}^-)$  & 20.15 & ($2f^{n}_{{5/2}}$~-~$3d^{p}_{{5/2}}$) & $-$0.97  \\
		& & 21.31 & ($2d^{n}_{{3/2}}$~-~$3p^{p}_{{3/2}}$) & $-$0.99  \\
		& & 22.47 & ($3p^{n}_{{3/2}}$~-~$3d^{p}_{{3/2}}$) &  0.98  \\
		& & 23.72 & ($1g^{n}_{{9/2}}$~-~$1h^{p}_{{9/2}}$) &  0.87  \\
		& &       & ($2d^{n}_{{5/2}}$~-~$2f^{p}_{{5/2}}$) &  0.45  \\
		& & 24.53 & ($2f^{n}_{{7/2}}$~-~$2g^{p}_{{7/2}}$) &  0.97  \\
		& &       & ($1g^{n}_{{9/2}}$~-~$1h^{p}_{{9/2}}$) &  $-$0.32  \\
		& & 25.14 & ($1h^{n}_{{11/2}}$~-~$1i^{p}_{{11/2}}$) & 0.96   \\
		& &       & ($2f^{n}_{{7/2}}$~-~$2g^{p}_{{7/2}}$) & 0.31   \\

		$^{214}$Pb$(0^+)$ & $^{214}$Bi$(2_{1}^-)$ &  19.32 & ($2f^{n}_{{5/2}}$~-~$3d^{p}_{{5/2}}$) & $-$0.99 \\
		&  & 19.83 & ($2d^{n}_{{3/2}}$~-~$3p^{p}_{{3/2}}$) & $-$0.98 \\
		&  & 21.18 & ($3p^{n}_{{3/2}}$~-~$3d^{p}_{{3/2}}$) & 0.98 \\
		&  & 22.46 & ($1g^{n}_{{9/2}}$~-~$1h^{p}_{{9/2}}$) & 0.86 \\
		&  & 23.07 & ($2f^{n}_{{7/2}}$~-~$2g^{p}_{{7/2}}$) & 0.97 \\
		&  & 23.68 & ($1h^{n}_{{11/2}}$~-~$1i^{p}_{{11/2}}$) & 0.96 \\
		\hline
	\end{tabular}
	\end{table}

	\begin{table}
	\centering
	\caption{Same as Table~\ref{tab3} but for selected Bi \textrightarrow Po transitions. }\label{tab4}
	\scriptsize
	\renewcommand{\arraystretch}{2.0}
	\renewcommand{\tabcolsep}{0.12cm}
	\renewcommand{\ULdepth}{6.0pt}
	\vspace{0.25cm}
	\begin{tabular}{ccccc}
		\hline			
		\multicolumn{1}{c}{Initial ($J_i^{\pi}$) } &
		\multicolumn{1}{c}{Final ($J_f^{\pi}$)}  &
		\multicolumn{1}{c}{~$\omega_{i}$ ($\mathrm MeV$)} &
		\multicolumn{1}{c}{~Structure} &
		\multicolumn{1}{c}{~$\psi_{np}^{i}$} \\
		\hline
		
		$^{210}$Bi$(1^-)$ & $^{210}$Po$(0_1^+)$ & 20.05 & ($3p^{n}_{{1/2}}$~-~$4s^{p}_{{1/2}}$) & 0.96 \\
		&  & 20.46 & ($2d^{n}_{{3/2}}$~-~$3p^{p}_{{3/2}}$) & $-$0.98 \\
		&  & 21.72 & ($1g^{n}_{{9/2}}$~-~$1h^{p}_{{9/2}}$) & 0.82 \\
		&  &  & ($2d^{n}_{{5/2}}$~-~$2f^{p}_{{5/2}}$) & 0.45 \\
		&  & 22.03 & ($3p^{n}_{{3/2}}$~-~$3d^{p}_{{3/2}}$) & 0.65 \\
		&  & 22.84 & ($2f^{n}_{{7/2}}$~-~$2g^{p}_{{7/2}}$) & 0.96 \\
		&  & 23.76 & ($1g^{n}_{{9/2}}$~-~$1h^{p}_{{9/2}}$)  & 0.74 \\
		&  & 24.57 & ($1h^{n}_{{11/2}}$~-~$1i^{p}_{{11/2}}$)  & 0.99\\
		
		$^{212}$Bi$(1^-)$ & $^{212}$Po$(0_1^+)$ & 20.74 & ($3p^{n}_{{1/2}}$~-~$4s^{p}_{{1/2}}$)  & $-$ 0.99 \\
		&  & 21.53 & ($3p^{n}_{{1/2}}$~-~$4s^{p}_{{1/2}}$) & $-$ 0.98\\
		&  & 22.82 & ($1g^{n}_{{9/2}}$~-~$1h^{p}_{{9/2}}$) & 0.98\\
		&  & 23.62 & ($2f^{n}_{{7/2}}$~-~$2g^{p}_{{7/2}}$) & $-$ 0.42\\
		&  & 23.91 & ($1h^{n}_{{11/2}}$~-~$1i^{p}_{{11/2}}$) & 0.94 \\
		&  & 24.87 & ($2g^{n}_{{7/2}}$~-~$3f^{p}_{{7/2}}$) & $-$ 0.98 \\
		&  &       & ($1h^{n}_{{11/2}}$~-~$1i^{p}_{{11/2}}$) & 0.19 \\
		
		$^{214}$Bi$(1^-)$ & $^{214}$Po$(0_1^+)$ & 19.72 & ($2f^{n}_{{5/2}}$~-~$3d^{p}_{{5/2}}$) & $-$ 0.99 \\
		&  &       & ($2g^{n}_{{7/2}}$~-~$1f^{p}_{{7/2}}$) & $-$ 0.23 \\
		&  & 20.81 & ($3p^{n}_{{3/2}}$~-~$3d^{p}_{{3/2}}$) & 0.97\\
		&  & 21.79 & ($3p^{n}_{{3/2}}$~-~$3d^{p}_{{3/2}}$) & 0.96 \\
		&  & 22.52 & ($1g^{n}_{{9/2}}$~-~$1h^{p}_{{9/2}}$) & 0.85 \\
		&  &       & ($2d^{n}_{{5/2}}$~-~$2f^{p}_{{5/2}}$) & 0.41 \\
		&  & 23.15 & ($2f^{n}_{{7/2}}$~-~$2g^{p}_{{7/2}}$) & 0.94 \\
		&  & 24.63 & ($1h^{n}_{{11/2}}$~-~$1i^{p}_{{11/2}}$) & 0.86 \\	
		&  & 25.57 & ($1h^{n}_{{11/2}}$~-~$1i^{p}_{{11/2}}$) & 0.97 \\	
		&  &       & ($2f^{n}_{{7/2}}$~-~$2g^{p}_{{7/2}}$) & 0.23 \\	
		&  &       & ($1g^{n}_{{9/2}}$~-~$1h^{p}_{{9/2}}$) & 0.17 \\		
		
		\hline
	\end{tabular}
	\end{table}	
	\clearpage
	\begin{table}
	\centering
	\caption{Comparison between calculated and measured half-lives  for  Pb \textrightarrow Bi transitions. The pn-QRPA(WS)$^{*}$ (pn-QRPA(WS)) values refer to our model calculation with (without) quenching factor. The experimental data and the SM I $/$ SM II values were taken from refs. \cite{Nudat} and \cite{Sha22}, respectively.}\label{tab5}
		\scriptsize
		\renewcommand{\arraystretch}{2.0}
		\renewcommand{\tabcolsep}{0.11cm}
		\renewcommand{\ULdepth}{6.0pt}
		\vspace{0.25cm}
		\begin{tabular}{ccccccc}
			\hline
			\multicolumn{2}{c}{\uline{~~~~~~~~~Transition~~~~~~~}~~~~~~} & 
			\multicolumn{5}{c}{\uline{~~~~~~~~~~~~~~~~~~~~~~~~~~~~~~~~~~~~~~~~~~~~~~~~ Half-lives (s) ~~~~~~~~~~~~~~~~~~~~~~~~~~~~~~~~~~~~~~~~~~~~~~~~~}} \\
			
			\multicolumn{1}{c}{Initial~~~~} & 
			\multicolumn{1}{c}{~~~~Final~~~~} & 
			\multicolumn{1}{c}{~Exp.~~} & 
			\multicolumn{1}{c}{~pn-QRPA(WS)} & 
			\multicolumn{1}{c}{~pn-QRPA(WS)$^{*}$} & 
			\multicolumn{1}{c}{~SM I} & 
			\multicolumn{1}{c}{~SM II ~} \\
			
			\hline
			
			\multicolumn{1}{l}{$^{210}$Pb$(0^+)$} 
			& \multicolumn{1}{c}{$^{210}$Bi} 
			& 7.00$\times10^8$ &  5.58$\times10^8$ & 6.58$\times10^8$ & 2.69$\times10^7$ & 4.19$\times10^7$\\
			
			\multicolumn{1}{l}{$^{211}$Pb$(9/2^+)$} 
			& \multicolumn{1}{c}{$^{211}$Bi} 
			& 2.16$\times10^3$ & 1.90$\times10^3$ & 2.42$\times10^3$ & 2.43$\times10^3$ & 2.58$\times10^3$ \\
			
			\multicolumn{1}{l}{$^{212}$Pb$(0^+)$} 
			& \multicolumn{1}{c}{$^{212}$Bi} 
			& 3.58$\times10^4$ &  2.56$\times10^4$ & 3.17$\times10^4$ & 1.46$\times10^4$ & 1.61$\times10^4$ \\
			
			\multicolumn{1}{l}{$^{213}$Pb$(9/2^+)$} 
			& \multicolumn{1}{c}{$^{213}$Bi} 
			& 6.12$\times10^2$ & 8.41$\times10^2$ & 9.73$\times10^2$ & 9.73$\times10^2$ & 1.11$\times10^2$ \\
			
			\multicolumn{1}{l}{$^{214}$Pb$(0^+)$} 
			& \multicolumn{1}{c}{$^{214}$Bi} 
			& 1.62$\times10^3$ & 9.39$\times10^3$ & 1.28$\times10^3$ & 6.33$\times10^2$ & 2.02$\times10^2$ \\
			
			\multicolumn{1}{l}{$^{215}$Pb$(9/2^+)$} 
			& \multicolumn{1}{c}{$^{215}$Bi} 
			& 1.47$\times10^2$ & 4.32$\times10^2$ & 6.03$\times10^2$ & 6.75$\times10^2$ & 8.13$\times10^2$ \\
			
			\hline
			
	\end{tabular}
	\end{table}
	
	\begin{table}
		\centering
		\caption{Same as  Table~\ref{tab5} but for  Bi \textrightarrow Po transitions.}\label{tab6}
		\scriptsize
		\renewcommand{\arraystretch}{2.0}
		\renewcommand{\tabcolsep}{0.11cm}
		\renewcommand{\ULdepth}{6.0pt}
		\vspace{0.25cm}
		\begin{tabular}{ccccccc}
			\hline
			\multicolumn{2}{c}{\uline{~~~~~~~~~Transition~~~~~~~}~~~~~~} & 
			\multicolumn{5}{c}{\uline{~~~~~~~~~~~~~~~~~~~~~~~~~~~~~~~~~~~~~~~~~~~~~~~~ Half-lives (s)
					~~~~~~~~~~~~~~~~~~~~~~~~~~~~~~~~~~~~~~~~~~~~~~~~~~~~ }} \\
			
			\multicolumn{1}{c}{Initial~~~~} & 
			\multicolumn{1}{c}{~~~~Final~~~~} & 
			\multicolumn{1}{c}{~Exp.~~} & 
			\multicolumn{1}{c}{~pn-QRPA(WS)} & 
			\multicolumn{1}{c}{~pn-QRPA(WS)$^{*}$} & 
			\multicolumn{1}{c}{~SM I~} & 
			\multicolumn{1}{c}{~SM II~} \\
			
			\hline
			
			\multicolumn{1}{l}{$^{210}$Bi$(1^-)$} 
			& \multicolumn{1}{c}{$^{210}$Po} 
			& 4.33$\times10^5$ & 3.26$\times10^5$ & 4.73$\times10^5$ & 2.35$\times10^5$ & 6.11$\times10^6$\\
			
			\multicolumn{1}{l}{$^{211}$Bi$(9/2^-)$} 
			& \multicolumn{1}{c}{$^{211}$Po} 
			& 1.28$\times10^2$ & 9.13$\times10^1$ & 1.14$\times10^2$ & 1.65$\times10^2$ & 1.64$\times10^2$ \\
			
			\multicolumn{1}{l}{$^{212}$Bi$(1^-)$} 
			& \multicolumn{1}{c}{$^{212}$Po} 
			& 3.63$\times10^3$ & 3.32$\times10^3$ & 5.24$\times10^3$ & 4.77$\times10^4$ & 2.79$\times10^5$ \\
			
			\multicolumn{1}{l}{$^{213}$Bi$(9/2^-)$} 
			& \multicolumn{1}{c}{$^{213}$Po} 
			& 2.73$\times10^2$ & 2.22$\times10^2$ & 2.48$\times10^3$ & 1.62$\times10^3$ & 2.99$\times10^3$ \\
			
			\multicolumn{1}{l}{$^{214}$Bi$(1^-)$} 
			& \multicolumn{1}{c}{$^{214}$Po} 
			& 1.18$\times10^3$ & 7.96$\times10^2$ & 9.61$\times10^2$ & 5.33$\times10^2$ & 1.80$\times10^3$ \\
			
			\multicolumn{1}{l}{$^{215}$Bi$(9/2^-)$} 
			& \multicolumn{1}{c}{$^{215}$Po} 
			& 4.56$\times10^2$ & 3.03$\times10^2$ & 5.10$\times10^2$ & 2.04$\times10^2$ & 4.89$\times10^2$ \\
			
			\hline
			
		\end{tabular}
	\end{table}

	\begin{table}[!ht]
		\centering
		\caption{Accuracy of the model calculations compared to experimental data. For explanation of symbols see text.} \label{tab7}
		\scriptsize
		\renewcommand{\arraystretch}{2.0}
		\renewcommand{\tabcolsep}{0.11cm}
		\renewcommand{\ULdepth}{6.0pt}
		\vspace{0.25cm}
		\begin{tabular}{l|llll|l|llll}
			\hline 
			Nuclei & ~ ~ & ~ & ~ ~ & ~ & Nuclei & ~ ~ & ~ & ~ ~ & ~ \\ \hline
			~ & $X_{WS}$ & $X_{(WS)}$* & ~$X_{SM I}$ & $X_{SM II}$ & ~ & $X_{WS}$ & $X _{(WS)}$* & ~$X_{SM I}$ & $X_{SM II}$ \\ \hline
			$^{210}$Pb & 1.19 & 1.06 & ~25.05 & 16.45 & $^{210}$Bi & 1.33 & 1.09 & ~22.07 & 1.17 \\ 
			$^{211}$Pb & 1.13 & 1.11 & ~1.12 & 1.19 & $^{211}$Bi & 1.4 & 1.11 & ~1.28 & 1.27 \\ 
			$^{212}$Pb & 1.48 & 1.20 & ~2.61 & 2.37 & $^{212}$Bi & 1.09 & 1.44 & ~1.31 & 7.69 \\ 
			$^{213}$Pb & 1.37 & 1.59 & ~1.59 & 1.82 & $^{213}$Bi & 1.23 & 1.10 & ~1.68 & 1.09 \\ 
			$^{214}$Pb & 1.72 & 1.26 & ~2.56 & 1.24 & $^{214}$Bi & 1.48 & 1.23 & ~2.21 & 1.52 \\ 
			$^{215}$Pb & 2.93 & 4.10 & ~4.59 & 5.53 & $^{215}$Bi & 1.50 & 1.12 & ~2.22 & 1.07 \\ 
	\hline
			$\bar{X}$ & 1.64 & 1.72 &  6.25 & 4.77 & $\bar{X}$ & 1.34 &  1.18 & 5.13 & 2.30\\
	\end{tabular}
	\end{table}
	\clearpage
	\begin{table}
	\centering
	\caption{Combined $\beta$-decay and positron capture rates for $^{210,211}$Pb in stellar matter using the pn-QRPA(N) model in units of $s^{-1}$. Density has units of  $g.cm^{-3}$ and $T_9$ is the core temperature in units of $10^9 K$. The last two columns show the percentage contribution of allowed and U1F rates.} \label{tab8}
	\scriptsize
	
	\renewcommand{\arraystretch}{2.0}
	\renewcommand{\tabcolsep}{0.12cm}
	\renewcommand{\ULdepth}{6.0pt}
	\vspace{0.25cm}
	\centering
	
	\begin{tabular}{c|c|c|ccc|cc}
		\multicolumn{1}{c|}{Element} & 
		\multicolumn{1}{c|}{Density } & 
		\multicolumn{1}{c|}{$T_9$} & 
		\multicolumn{3}{c|}{$\beta$-decay and positron capture rates} &
		\multicolumn{1}{c}{~~\% contribution of GT and U1F}  & \\ 
		\hline
		\multicolumn{1}{c|}{~ ~} & 
		\multicolumn{1}{c|}{~ ~} & 
		\multicolumn{1}{c|}{~ ~} & 
		\multicolumn{1}{c}{GT} & 
		\multicolumn{1}{c}{U1F} & 
		\multicolumn{1}{c|}{Total}  &
		\multicolumn{1}{c}{GT\%} & 
		\multicolumn{1}{c}{U1F\%} \\ 
		\hline
		$^{210}$Pb & $10^1$ & 1.5 & 3.85E-07 & 7.19E-07 & 1.10E-06 & 34.87 & 65.13 \\
		~ & ~ & 10 & 1.86E-01 & 7.01E+02 & 7.01E+02 & 0.03 & 99.97 \\
		~ & ~ & 20 & 1.34E+02 & 5.95E+03 & 6.08E+03 & 2.20 & 97.80 \\
		~ & ~ & 30 & 3.65E+03 & 1.65E+04 & 2.02E+04 & 18.11 & 81.89  \\
		~ & $10^3$ & 1.5 & 3.76E-07 & 7.19E-07 & 1.10E-06 & 34.34 & 65.66 \\
		~ & ~ & 10 & 1.86E-01 & 7.01E+02 & 7.01E+02 & 0.03 & 99.97 \\
		~ & ~ & 20 & 1.34E+02 & 5.95E+03 & 6.08E+03 & 2.20 & 97.80 \\ 
		~ & ~ & 30 & 3.66E+03 & 1.65E+04 & 2.02E+04 & 18.15 & 81.85 \\ 
		~ & $10^7$ & 1.5 & 1.03E-09 & 5.05E-07 & 5.06E-07 & 0.20 & 99.80 \\ 
		~ & ~ & 10 & 1.48E-01 & 6.80E+02 & 6.80E+02 & 0.02 & 99.98 \\
		~ & ~ & 20 & 1.31E+02 & 5.88E+03 & 6.01E+03 & 21.18 & 97.82 \\ 
		~ & ~ & 30 & 3.63E+03 & 1.64E+04 & 2.00E+04 & 18.12 & 81.88 \\ 
		~ & $10^{11}$ & 1.5 & 3.53E-86 & 1.49E-77 & 1.49E-77 & 0.00 & 100.00 \\ 
		~ & ~ & 10 & 1.81E-13 & 2.11E-08 & 2.11E-08 & 0.00 & 100.00 \\ 
		~ & ~ & 20 & 1.58E-04 & 3.68E-02 & 3.70E-02 & 0.43 & 99.57 \\
		~ & ~ & 30 & 4.99E-01 & 4.88E+00 & 5.38E+00 & 9.28 & 90.72 \\ \hline
		$^{211}$Pb & $10^1$ & 1.5 & 8.20E-03 & 6.46E-02 & 7.28E-02 & 11.26 & 88.74 \\
		~ & ~ & 10 & 1.19E-01 & 3.48E+01 & 3.49E+01 & 0.34 & 99.66 \\
		~ & ~ & 20 & 5.22E+01 & 5.95E+02 & 6.47E+02 & 8.07 & 91.93 \\
		~ & ~ & 30 & 1.89E+03 & 3.40E+03 & 5.29E+03 & 35.73 & 64.27 \\ 
		~ & $10^3$ & 1.5 & 8.20E-03 & 6.46E-02 & 7.28E-02 & 11.26 & 88.74 \\ 
		~ & ~ & 10 & 1.19E-01 & 3.48E+01 & 3.49E+01 & 0.34 & 99.66 \\ 
		~ & ~ & 20 & 5.24E+01 & 5.96E+02 & 6.48E+02 & 8.08 & 91.92 \\
		~ & ~ & 30 & 1.89E+03 & 3.41E+03 & 5.30E+03 & 35.66 & 64.34 \\ 
		~ & $10^7$ & 1.5 & 2.85E-03 & 4.49E-03 & 7.34E-03 & 38.83 & 61.17 \\
		~ & ~ & 10 & 9.53E-02 & 2.99E+01 & 3.00E+01 & 0.32 & 99.68 \\
		~ & ~ & 20 & 5.11E+01 & 5.80E+02 & 6.31E+02 & 8.10 & 91.90 \\ 
		~ & ~ & 30 & 1.88E+03 & 3.39E+03 & 5.27E+03 & 35.67 & 64.33 \\ 
		~ & $10^{11}$ & 1.5 & 5.93E-79 & 4.05E-78 & 4.64E-78 & 12.77 & 87.23 \\
		~ & ~ & 10 & 1.28E-13 & 9.33E-11 & 9.34E-11 & 0.14 & 99.86 \\ 
		~ & ~ & 20 & 6.19E-05 & 7.74E-04 & 8.36E-04 & 7.41 & 92.59 \\ 
		~ & ~ & 30 & 2.58E-01 & 4.85E-01 & 7.43E-01 & 34.72 & 65.28 \\
		\hline
	\end{tabular}
	\end{table}
	\begin{table}
	\centering
	\caption{Same as Table~\ref{tab8} but for $^{212,213}$Pb. } \label{tab9}
	\scriptsize
	
	\renewcommand{\arraystretch}{2.0}
	\renewcommand{\tabcolsep}{0.12cm}
	\renewcommand{\ULdepth}{6.0pt}
	\vspace{0.25cm}
	\centering
	
	\begin{tabular}{c|c|c|ccc|cc}
		\multicolumn{1}{c|}{Element} & 
\multicolumn{1}{c|}{Density } & 
\multicolumn{1}{c|}{$T_9$} & 
\multicolumn{3}{c|}{$\beta$-decay and positron capture rates} &
\multicolumn{1}{c}{~~\% contribution of GT and U1F}  & \\ 
\hline
\multicolumn{1}{c|}{~ ~} & 
\multicolumn{1}{c|}{~ ~} & 
\multicolumn{1}{c|}{~ ~} & 
\multicolumn{1}{c}{GT} & 
\multicolumn{1}{c}{U1F} & 
\multicolumn{1}{c|}{Total}  &
\multicolumn{1}{c}{GT\%} & 
\multicolumn{1}{c}{U1F\%} \\ 
\hline
		$^{212}$Pb & $10^1$ & 1.5  & 4.22E-07 & 2.76E-02 & 2.76E-02 & 0.00 & 100.00 \\ 
		~ & ~ & 10 & 2.57E-01 & 1.06E+03 & 1.06E+03 & 0.02 & 99.98 \\ 
		~ & ~ & 20 & 1.64E+02 & 5.62E+03 & 5.78E+03 & 2.84 & 97.16 \\ 
		~ & ~ & 30 & 6.58E+03 & 1.47E+04 & 2.13E+04 & 30.92 & 69.08 \\ 
		~ & $10^3$ & 1.5 & 4.09E-07 & 2.76E-02 & 2.76E-02 & 0.00 & 100.00 \\ 
		~ & ~ & 10 & 2.58E-01 & 1.06E+03 & 1.06E+03 & 0.02 & 99.98 \\
		~ & ~ & 20 & 1.65E+02 & 5.63E+03 & 5.80E+03 & 2.85 & 97.15 \\
		~ & ~ & 30 & 6.59E+02 & 1.48E+04 & 1.55E+04 & 4.26 & 95.74 \\ 
		~ & $10^7$ & 1.5 & 2.49E-10 & 8.91E-04 & 8.91E-04 & 0.00 & 100.00 \\
		~ & ~ & 10 & 2.06E-01 & 1.03E+03 & 1.03E+03 & 0.02 & 99.98 \\
		~ & ~ & 20 & 1.60E+02 & 5.55E+03 & 5.71E+03 & 2.80 & 97.20 \\
		~ & ~ & 30 & 6.55E+03 & 1.47E+04 & 2.13E+04 & 30.82 & 69.18 \\ 
		~ & $10^{11}$ & 1.5 & 8.45E-87 & 9.27E-76 & 9.27E-76 & 0.00 & 100.00 \\ 
		~ & ~ & 10 & 2.52E-13 & 2.25E-08 & 2.25E-08 & 0.00 & 100.00 \\ 
		~ & ~ & 20 & 1.95E-04 & 2.67E-02 & 2.69E-02 & 0.73 & 99.27 \\ 
		~ & ~ & 30 & 8.99E-01 & 3.57E+00 & 4.47E+00 & 20.12 & 79.88 \\ \hline
		$^{213}$Pb & $10^1$ & 1.5 & 2.73E-01 & 2.64E-01 & 5.37E-01 & 50.84 & 49.16 \\ 
		~ & ~ & 10 & 4.44E+00 & 3.64E+02 & 3.68E+02 & 1.21 & 98.79 \\
		~ & ~ & 20 & 9.94E+01 & 9.55E+02 & 1.05E+03 & 9.43 & 90.57 \\
		~ & ~ & 30 & 2.69E+03 & 5.34E+03 & 8.03E+03 & 33.50 & 66.50 \\ 
		~ & $10^3$ & 1.5 & 2.73E-01 & 2.64E-01 & 5.37E-01 & 50.84 & 49.16 \\ 
		~ & ~ & 10 & 4.44E+00 & 3.64E+02 & 3.68E+02 & 1.21 & 98.79 \\ 
		~ & ~ & 20 & 9.99E+01 & 9.56E+02 & 1.06E+03 & 9.46 & 90.54 \\
		~ & ~ & 30 & 2.70E+03 & 5.35E+03 & 8.05E+03 & 33.54 & 66.46 \\ 
		~ & $10^7$ & 1.5 & 1.29E-01 & 7.62E-02 & 2.05E-01 & 62.87 & 37.13 \\
		~ & ~ & 10 & 4.15E+00 & 3.52E+02 & 3.56E+02 & 11.7 & 98.83 \\
		~ & ~ & 20 & 9.72E+01 & 9.41E+02 & 1.04E+03 & 9.36 & 90.64 \\
		~ & ~ & 30 & 2.68E+03 & 5.30E+03 & 7.98E+03 & 33.58 & 66.42 \\ 
		~ & $10^{11}$ & 1.5 & 1.85E-75 & 1.42E-72 & 1.42E-72 & 0.13 & 99.87 \\
		~ & ~ & 10 & 4.92E-11 & 6.04E-09 & 6.09E-09 & 0.81 & 99.19 \\
		~ & ~ & 20 & 1.34E-04 & 3.55E-03 & 3.68E-03 & 3.64 & 96.36 \\ 
		~ & ~ & 30 & 3.70E-01 & 9.18E-01 & 1.29E+00 & 28.73 & 71.27 \\
		\hline
	\end{tabular}
	\end{table}
	\begin{table}
	\centering
	\caption{Same as Table~\ref{tab8} but for $^{214,215}$Pb. } \label{tab10}
	\scriptsize
	
	\renewcommand{\arraystretch}{2.0}
	\renewcommand{\tabcolsep}{0.12cm}
	\renewcommand{\ULdepth}{6.0pt}
	\vspace{0.25cm}
	\centering
	
	\begin{tabular}{c|c|c|ccc|cc}
		\multicolumn{1}{c|}{Element} & 
\multicolumn{1}{c|}{Density } & 
\multicolumn{1}{c|}{$T_9$} & 
\multicolumn{3}{c|}{$\beta$-decay and positron capture rates} &
\multicolumn{1}{c}{~~\% contribution of GT and U1F}  & \\ 
\hline
\multicolumn{1}{c|}{~ ~} & 
\multicolumn{1}{c|}{~ ~} & 
\multicolumn{1}{c|}{~ ~} & 
\multicolumn{1}{c}{GT} & 
\multicolumn{1}{c}{U1F} & 
\multicolumn{1}{c|}{Total}  &
\multicolumn{1}{c}{GT\%} & 
\multicolumn{1}{c}{U1F\%} \\ 
\hline
		$^{214}$Pb & $10^1$ & 1.5 & 5.75E-07 & 2.02E-02 & 2.02E-02 & 0.00 & 100.00 \\
		~ & ~ & 10 & 3.94E-01 & 8.39E+02 & 8.39E+02 & 0.05 & 99.95 \\ 
		~ & ~ & 20 & 1.90E+02 & 2.59E+03 & 2.78E+03 & 6.83 & 93.17 \\
		~ & ~ & 30 & 4.69E+03 & 7.15E+03 & 1.18E+04 & 39.61 & 60.39 \\ 
		~ & $10^3$ & 1.5 & 5.56E-07 & 2.02E-02 & 2.02E-02 & 0.00 & 100.00 \\
		~ & ~ & 10 & 3.95E-01 & 8.39E+02 & 8.39E+02 & 0.05 & 99.95 \\
		~ & ~ & 20 & 1.90E+02 & 2.60E+03 & 2.79E+03 & 6.81 & 93.19 \\
		~ & ~ & 30 & 4.70E+03 & 7.17E+03 & 1.19E+04 & 39.60 & 60.40 \\ 
		~ & $10^7$ & 1.5 & 1.35E-10 & 1.31E-03 & 1.31E-03 & 0.00 & 100.00 \\ 
		~ & ~ & 10 & 3.15E-01 & 8.11E+02 & 8.11E+02 & 0.04 & 99.96 \\
		~ & ~ & 20 & 1.85E+02 & 2.56E+03 & 2.75E+03 & 6.74 & 93.26 \\
		~ & ~ & 30 & 4.67E+03 & 7.11E+03 & 1.18E+04 & 39.64 & 60.36 \\ 
		~ & $10^{11}$ & 1.5 & 4.60E-87 & 3.78E-75 & 3.78E-75 & 0.00 & 100.00 \\
		~ & ~ & 10 & 3.86E-13 & 1.31E-08 & 1.31E-08 & 0.00 & 100.00 \\ 
		~ & ~ & 20 & 2.25E-04 & 9.91E-03 & 1.01E-02 & 2.22 & 97.78 \\
		~ & ~ & 30 & 6.41E-01 & 1.48E+00 & 2.12E+00 & 30.22 & 69.78 \\ \hline
		$^{215}$Pb & $10^1$ & 1.5 & 1.29E+00 & 8.85E+01 & 8.98E+01 & 1.44 & 98.56 \\
		~ & ~ & 10 & 8.63E+00 & 2.73E+02 & 2.82E+02 & 3.06 & 96.94 \\ 
		~ & ~ & 20 & 8.77E+01 & 5.33E+02 & 6.21E+02 & 14.13 & 85.87 \\
		~ & ~ & 30 & 1.27E+03 & 2.18E+03 & 3.45E+03 & 36.81 & 63.19 \\ 
		~ & $10^3$ & 1.5 & 1.29E+00 & 8.85E+01 & 8.98E+01 & 1.44 & 98.56 \\
		~ & ~ & 10 & 8.62E+00 & 2.73E+02 & 2.82E+02 & 3.06 & 96.94 \\
		~ & ~ & 20 & 8.79E+01 & 5.33E+02 & 6.21E+02 & 14.16 & 85.84 \\ 
		~ & ~ & 30 & 1.27E+03 & 2.19E+03 & 3.46E+03 & 36.71 & 63.29 \\ 
		~ & $10^7$ & 1.5 & 9.75E-01 & 6.05E+01 & 6.15E+01 & 1.59 & 98.41 \\ 
		~ & ~ & 10 & 8.22E+00 & 2.64E+02 & 2.72E+02 & 3.02 & 96.98 \\
		~ & ~ & 20 & 8.56E+01 & 5.25E+02 & 6.11E+02 & 14.02 & 85.98 \\ 
		~ & ~ & 30 & 1.26E+03 & 2.17E+03 & 3.43E+03 & 36.73 & 63.27 \\ 
		~ & $10^{11}$ & 1.5 & 5.94E-73 & 2.00E-70 & 2.01E-70 & 0.30 & 99.70 \\ 
		~ & ~ & 10 & 1.38E-10 & 4.50E-09 & 4.64E-09 & 2.98 & 97.02 \\ 
		~ & ~ & 20 & 1.35E-04 & 1.88E-03 & 2.01E-03 & 6.70 & 93.30 \\ 
		~ & ~ & 30 & 1.76E-01 & 3.76E-01 & 5.52E-01 & 31.88 & 68.12 \\ 
		\hline 
	\end{tabular}
	\end{table}
	\begin{table}[!ht]
	\centering
	\caption{Same as Table~\ref{tab8} but for $^{210,211}$Bi.} \label{tab11}
	\scriptsize
	\renewcommand{\arraystretch}{2.0}
	\renewcommand{\tabcolsep}{0.12cm}
	\renewcommand{\ULdepth}{6.0pt}
	\vspace{0.25cm}
	\begin{tabular}{c|c|c|ccc|cc}
		\multicolumn{1}{c|}{Element} & 
\multicolumn{1}{c|}{Density } & 
\multicolumn{1}{c|}{$T_9$} & 
\multicolumn{3}{c|}{$\beta$-decay and positron capture rates} &
\multicolumn{1}{c}{~~\% contribution of GT and U1F}  & \\ 
\hline
\multicolumn{1}{c|}{~ ~} & 
\multicolumn{1}{c|}{~ ~} & 
\multicolumn{1}{c|}{~ ~} & 
\multicolumn{1}{c}{GT} & 
\multicolumn{1}{c}{U1F} & 
\multicolumn{1}{c|}{Total}  &
\multicolumn{1}{c}{GT\%} & 
\multicolumn{1}{c}{U1F\%} \\ 
\hline
		
		$^{210}$Bi & $10^1$ & 1.5 & 7.99E-07 & 1.06E-06 & 1.86E-06 & 42.98 & 57.02 \\ 
		~ & ~ & 10 & 3.53E-02 & 2.14E-03 & 3.74E-02 & 94.28 & 5.72 \\
		~ & ~ & 20 & 1.39E+02 & 5.62E+00 & 1.45E+02 & 96.11 & 3.89 \\ 
		~ & ~ & 30 & 1.06E+04 & 2.82E+02 & 1.09E+04 & 97.41 & 2.59 \\ 
		~ & $10^3$ & 1.5 & 7.99E-07 & 1.01E-06 & 1.81E-06 & 44.17 & 55.83 \\ 
		~ & ~ & 10 & 3.53E-02 & 2.14E-03 & 3.74E-02 & 94.28 & 5.72 \\ 
		~ & ~ & 20 & 1.40E+02 & 5.62E+00 & 1.46E+02 & 96.14 & 3.86 \\ 
		~ & ~ & 30 & 1.06E+04 & 2.82E+02 & 1.09E+04 & 97.41 & 2.59 \\ 
		~ & $10^7$ & 1.5 & 5.40E-08 & 5.38E-08 & 1.08E-07 & 50.09 & 49.91 \\
		~ & ~ & 10 & 2.81E-02 & 1.71E-03 & 2.98E-02 & 94.26 & 5.74 \\ 
		~ & ~ & 20 & 1.36E+02 & 5.47E+00 & 1.41E+02 & 96.13 & 3.87 \\
		~ & ~ & 30 & 1.06E+04 & 2.80E+02 & 1.09E+04 & 97.43 & 2.57 \\ 
		~ & $10^{11}$ & 1.5 & 2.54E-84 & 2.36E-84 & 4.90E-84 & 51.84 & 48.16 \\
		~ & ~ & 10 & 3.44E-14 & 2.09E-15 & 3.65E-14 & 94.27 & 5.73 \\ 
		~ & ~ & 20 & 1.65E-04 & 6.64E-06 & 1.72E-04 & 96.13 & 3.87 \\ 
		~ & ~ & 30 & 1.45E+00 & 3.85E-02 & 1.49E+00 & 97.41 & 2.59 \\ \hline
		$^{211}$Bi & $10^1$ & 1.5 & 8.38E-08 & 3.63E-01 & 3.63E-01 & 0.00 & 100.00 \\
		~ & ~ & 10 & 2.94E-02 & 2.63E+00 & 2.66E+00 & 1.11\ & 98.89 \\
		~ & ~ & 20 & 1.45E+02 & 3.18E+01 & 1.77E+02 & 82.01 & 17.99 \\ 
		~ & ~ & 30 & 1.16E+04 & 4.20E+02 & 1.20E+04 & 96.51 & 3.49 \\ 
		~ & $10^3$ & 1.5 & 8.36E-08 & 3.62E-01 & 3.62E-01 & 0.00 & 100.00 \\
		~ & ~ & 10 & 2.95E-02 & 2.63E+00 & 2.66E+00 & 1.11 & 98.89 \\
		~ & ~ & 20 & 1.45E+02 & 3.19E+01 & 1.77E+02 & 81.97 & 18.03 \\ 
		~ & ~ & 30 & 1.16E+04 & 4.21E+02 & 1.20E+04 & 96.50 & 3.50 \\ 
		~ & $10^7$ & 1.5 & 5.60E-09 & 7.29E-03 & 7.29E-03 & 0.00 & 100.00 \\
		~ & ~ & 10 & 2.35E-02 & 2.12E+00 & 2.14E+00 & 1.10 & 98.90 \\ 
		~ & ~ & 20 & 1.41E+02 & 3.10E+01 & 1.72E+02 & 81.98 & 18.02 \\
		~ & ~ & 30 & 1.15E+04 & 4.18E+02 & 1.19E+04 & 96.49 & 3.51 \\ 
		~ & $10^{11}$ & 1.5 & 3.92E-85 & 2.58E-79 & 2.58E-79 & 0.00 & 100.00 \\ 
		~ & ~ & 10 & 2.88E-14 & 2.82E-12 & 2.85E-12 & 1.01 & 98.99 \\ 
		~ & ~ & 20 & 1.71E-04 & 3.88E-05 & 2.10E-04 & 81.51 & 18.49 \\
		~ & ~ & 30 & 1.58E+00 & 5.78E-02 & 1.64E+00 & 96.47 & 3.53 \\  				
		\hline
	\end{tabular}
	\end{table}
	\begin{table}[!ht]
	\centering
	\caption{Same as Table~\ref{tab8} but for $^{212,213}$Bi.} \label{tab12}
	\scriptsize
	\renewcommand{\arraystretch}{2.0}
	\renewcommand{\tabcolsep}{0.12cm}
	\renewcommand{\ULdepth}{6.0pt}
	\vspace{0.25cm}
	\begin{tabular}{c|c|c|ccc|cc}
		\multicolumn{1}{c|}{Element} & 
\multicolumn{1}{c|}{Density } & 
\multicolumn{1}{c|}{$T_9$} & 
\multicolumn{3}{c|}{$\beta$-decay and positron capture rates} &
\multicolumn{1}{c}{~~\% contribution of GT and U1F}  & \\ 
\hline
\multicolumn{1}{c|}{~ ~} & 
\multicolumn{1}{c|}{~ ~} & 
\multicolumn{1}{c|}{~ ~} & 
\multicolumn{1}{c}{GT} & 
\multicolumn{1}{c}{U1F} & 
\multicolumn{1}{c|}{Total}  &
\multicolumn{1}{c}{GT\%} & 
\multicolumn{1}{c}{U1F\%} \\ 
\hline
		$^{212}$Bi & $10^1$ & 1.5 & 6.46E-06 & 4.66E-06 & 1.11E-05 & 58.09 & 41.91 \\ 
		~ & ~ & 10 & 5.97E-02 & 4.03E-03 & 6.37E-02 & 93.68 & 6.32 \\ 
		~ & ~ & 20 & 1.87E+02 & 7.91E+00 & 1.95E+02 & 95.94 & 4.06 \\ 
		~ & ~ & 30 & 1.34E+04 & 3.63E+02 & 1.38E+04 & 97.36 & 2.64 \\ 
		~ & $10^3$ & 1.5 & 6.46E-06 & 4.66E-06 & 1.11E-05 & 58.09 & 41.91 \\ 
		~ & ~ & 10 & 5.98E-02 & 4.03E-03 & 6.38E-02 & 93.69 & 6.31 \\
		~ & ~ & 20 & 1.87E+02 & 7.93E+00 & 1.95E+02 & 95.93 & 4.07 \\
		~ & ~ & 30 & 1.34E+04 & 3.64E+02 & 1.38E+04 & 97.36 & 2.64 \\ 
		~ & $10^7$ & 1.5 & 3.89E-06 & 2.07E-06 & 5.96E-06 & 65.27 & 34.73 \\
		~ & ~ & 10 & 4.76E-02 & 3.21E-03 & 5.08E-02 & 93.68 & 6.32 \\
		~ & ~ & 20 & 1.82E+02 & 7.71E+00 & 1.90E+02 & 95.94 & 4.06 \\ 
		~ & ~ & 30 & 1.33E+04 & 3.61E+02 & 1.37E+04 & 97.36 & 2.64 \\ 
		~ & $10^{11}$ & 1.5 & 1.23E-80 & 4.45E-81 & 1.68E-80 & 73.43 & 26.57 \\
		~ & ~ & 10 & 5.84E-14 & 3.93E-15 & 6.23E-14 & 93.69 & 6.31 \\
		~ & ~ & 20 & 2.21E-04 & 9.35E-06 & 2.30E-04 & 95.94 & 4.06 \\
		~ & ~ & 30 & 1.83E+00 & 4.95E-02 & 1.88E+00 & 97.37 & 2.63\\ \hline
		$^{213}$Bi & $10^1$ & 1.5 & 1.59E-06 & 3.22E-02 & 3.22E-02 & 0.00 & 100.00 \\ 
		~ & ~ & 10 & 7.45E-02 & 2.33E+00 & 2.40E+00 & 3.10 & 96.90 \\
		~ & ~ & 20 & 2.19E+02 & 4.12E+01 & 2.60E+02 & 84.17 & 15.83 \\
		~ & ~ & 30 & 1.51E+04 & 5.69E+02 & 1.57E+04 & 96.37 & 3.63 \\ 
		~ & $10^3$ & 1.5 & 1.59E-06 & 3.22E-02 & 3.22E-02 & 0.00 & 100.00 \\ 
		~ & ~ & 10 & 7.45E-02 & 2.33E+00 & 2.40E+00 & 3.10 & 96.90 \\
		~ & ~ & 20 & 2.19E+02 & 4.12E+01 & 2.60E+02 & 84.17 & 15.83 \\
		~ & ~ & 30 & 1.52E+04 & 5.69E+02 & 1.58E+04 & 96.39 & 3.61 \\ 
		~ & $10^7$ & 1.5 & 3.33E-07 & 1.57E-03 & 1.57E-03 & 0.02 & 99.98 \\ 
		~ & ~ & 10 & 5.94E-02 & 1.87E+00 & 1.93E+00 & 3.08 & 96.92 \\
		~ & ~ & 20 & 2.13E+02 & 4.01E+01 & 2.53E+02 & 84.16 & 15.84 \\
		~ & ~ & 30 & 1.51E+04 & 5.65E+02 & 1.57E+04 & 96.39 & 3.61 \\ 
		~ & $10^{11}$ & 1.5 & 4.38E-83 & 7.06E-80 & 7.06E-80 & 0.06 & 99.94 \\ 
		~ & ~ & 10 & 7.28E-14 & 2.35E-12 & 2.42E-12 & 3.00 & 97.00 \\ 
		~ & ~ & 20 & 2.59E-04 & 4.98E-05 & 3.09E-04 & 83.87 & 16.13 \\
		~ & ~ & 30 & 2.07E+00 & 7.82E-02 & 2.15E+00 & 96.36 & 3.64 \\ 				
		\hline
	\end{tabular}
	\end{table}	
	\begin{table}[!ht]
	\centering
	\caption{Same as Table~\ref{tab8} but for $^{214,215}$Bi.} \label{tab13}
	\scriptsize
	\renewcommand{\arraystretch}{2.0}
	\renewcommand{\tabcolsep}{0.12cm}
	\renewcommand{\ULdepth}{6.0pt}
	\vspace{0.25cm}
	\begin{tabular}{c|c|c|ccc|cc}
		\multicolumn{1}{c|}{Element} & 
\multicolumn{1}{c|}{Density } & 
\multicolumn{1}{c|}{$T_9$} & 
\multicolumn{3}{c|}{$\beta$-decay and positron capture rates} &
\multicolumn{1}{c}{~~\% contribution of GT and U1F}  & \\ 
\hline
\multicolumn{1}{c|}{~ ~} & 
\multicolumn{1}{c|}{~ ~} & 
\multicolumn{1}{c|}{~ ~} & 
\multicolumn{1}{c}{GT} & 
\multicolumn{1}{c}{U1F} & 
\multicolumn{1}{c|}{Total}  &
\multicolumn{1}{c}{GT\%} & 
\multicolumn{1}{c}{U1F\%} \\ 
\hline
		$^{214}$Bi & $10^1$ & 1.5 & 1.72E-04 & 8.11E-05 & 2.53E-04 & 67.96 & 32.04 \\
		~ & ~ & 10 & 5.37E-02 & 4.13E-03 & 5.78E-02 & 92.86 & 7.14 \\
		~ & ~ & 20 & 3.52E+01 & 1.57E+00 & 3.68E+01 & 95.73 & 4.27 \\
		~ & ~ & 30 & 1.45E+03 & 4.04E+01 & 1.49E+03 & 97.29 & 2.71 \\ 
		~ & $10^3$ & 1.5 & 1.72E-04 & 8.11E-05 & 2.53E-04 & 67.96 & 32.04 \\
		~ & ~ & 10 & 5.39E-02 & 4.14E-03 & 5.80E-02 & 92.87 & 7.13 \\
		~ & ~ & 20 & 3.53E+01 & 1.58E+00 & 3.69E+01 & 95.72 & 4.28 \\
		~ & ~ & 30 & 1.45E+03 & 4.04E+01 & 1.49E+03 & 97.29 & 2.71 \\ 
		~ & $10^7$ & 1.5 & 1.46E-04 & 5.61E-05 & 2.02E-04 & 72.24 & 27.76 \\
		~ & ~ & 10 & 4.29E-02 & 3.30E-03 & 4.62E-02 & 92.86 & 7.14 \\ 
		~ & ~ & 20 & 3.44E+01 & 1.53E+00 & 3.59E+01 & 95.74 & 4.26 \\
		~ & ~ & 30 & 1.44E+03 & 4.01E+01 & 1.48E+03 & 97.29 & 2.71 \\ 
		~ & $10^{11}$ & 1.5 & 2.36E-76 & 4.94E-77 & 2.85E-76 & 82.69 & 17.31 \\
		~ & ~ & 10 & 5.30E-14 & 4.18E-15 & 5.72E-14 & 92.69 & 7.31 \\
		~ & ~ & 20 & 4.17E-05 & 1.86E-06 & 4.36E-05 & 95.73 & 4.27 \\ 
		~ & ~ & 30 & 1.98E-01 & 5.51E-03 & 2.04E-01 & 97.29 & 2.71 \\ \hline
		$^{215}$Bi & $10^1$ & 1.5 & 1.08E-04 & 7.73E-05 & 1.85E-04 & 58.28 & 41.72 \\
		~ & ~ & 10 & 1.57E-01 & 1.31E-02 & 1.70E-01 & 92.30 & 7.70 \\
		~ & ~ & 20 & 2.76E+02 & 1.25E+01 & 2.89E+02 & 95.67 & 4.33 \\ 
		~ & ~ & 30 & 1.71E+04 & 4.76E+02 & 1.76E+04 & 97.29 & 2.71 \\ 
		~ & $10^3$ & 1.5 & 1.08E-04 & 7.73E-05 & 1.85E-04 & 58.28 & 41.72 \\ 
		~ & ~ & 10 & 1.57E-01 & 1.31E-02 & 1.70E-01 & 92.30 & 7.70 \\
		~ & ~ & 20 & 2.77E+02 & 1.25E+01 & 2.90E+02 & 95.68 & 4.32 \\
		~ & ~ & 30 & 1.71E+04 & 4.76E+02 & 1.76E+04 & 97.29 & 2.71 \\ 
		~ & $10^7$ & 1.5 & 6.61E-05 & 3.48E-05 & 1.01E-04 & 65.51 & 34.49 \\
		~ & ~ & 10 & 1.25E-01 & 1.04E-02 & 1.35E-01 & 92.32 & 7.68 \\
		~ & ~ & 20 & 2.69E+02 & 1.22E+01 & 2.81E+02 & 95.66 & 4.34 \\
		~ & ~ & 30 & 1.70E+04 & 4.73E+02 & 1.75E+04 & 97.29 & 2.71 \\ 
		~ & $10^{11}$ & 1.5 & 4.56E-79 & 1.42E-79 & 5.98E-79 & 76.25 & 23.75 \\
		~ & ~ & 10 & 1.54E-13 & 1.30E-14 & 1.67E-13 & 92.22 & 7.78 \\
		~ & ~ & 20 & 3.27E-04 & 1.48E-05 & 3.42E-04 & 95.67 & 4.33 \\ 
		~ & ~ & 30 & 2.33E+00 & 6.50E-02 & 2.40E+00 & 97.29\% & 2.71 \\ \hline	
	\end{tabular}
	\end{table}

%	\begin{figure}
%	\centering
%	\includegraphics[scale=0.5]{Pb.jpg}
%	\caption{Calculated half-lives of Pb isotopes. Experimental data was taken from ref.~\cite{Nudat}.} \label{fig1}
%	\label{Half live Pb}
%	\end{figure}
%	\begin{figure}
%	\centering
%	\includegraphics[scale=0.5]{Bi.jpg}
%	\caption{Same as Fig.1 but for Bi isotopes.} \label{fig2}
%	\label{Half live Bi}
%	\end{figure}
%	\clearpage
	\end{document}